\documentclass[aps,pra]{revtex4-1} 
\usepackage{graphicx}

\def\ket#1{|\,#1\,\rangle}
\def\bra#1{\langle\, #1\,|}
\def\braket#1#2{\langle\, #1\,|\,#2\,\rangle}

\def\opone{\leavevmode\hbox{\small1\kern-3.8pt\normalsize1}}

\newcommand{\beq}{\begin{equation}}
\newcommand{\eeq}{\end{equation}}
\newcommand{\ba}{\begin{eqnarray}}
\newcommand{\ea}{\end{eqnarray}}
\newcommand{\bea}{\begin{eqnarray}}
\newcommand{\eea}{\end{eqnarray}}
\newcommand{\bma}{\begin{subequations}}
\newcommand{\ema}{\end{subequations}}
\newcommand{\bwt}{\begin{widetext}}
\newcommand{\ewt}{\end{widetext}}
\def\abs#1{|\,#1\,|}

\begin{document}

\title{Temporal compression of quantum information-carrying photons using a photon-echo quantum memory approach}

\author{S. A. Moiseev}
\affiliation{Institute for Quantum Information Science, and Department of Physics \& Astronomy,
University of Calgary, Canada}
 \affiliation{Kazan Physical-Technical Institute of the Russian Academy of Sciences, Russia}
\author{W. Tittel}
\affiliation{Institute for Quantum Information Science, and Department of Physics \& Astronomy,
University of Calgary, Canada }

\begin{abstract}
We study quantum compression and decompression of light pulses
that carry quantum information using a photon-echo quantum
memory technique with controllable
inhomogeneous broadening of an isolated atomic absorption line. We
investigate media with differently broadened absorption profiles,
transverse and longitudinal, finding that the recall efficiency can be as large as unity and that the
quantum information encoded into the photonic qubits can remain unperturbed. Our results provide new insight into reversible light-atom interaction, and are interesting in view of future quantum communication networks,
where pulse compression and decompression may play an important
role to increase the qubit rate, or to map quantum information
from photonic carriers with large optical bandwidth into atomic memories with smaller bandwidth.
\end{abstract}


\maketitle

\section{Introduction}{\label{Introduction}}

As any communication system, quantum communication relies on preparing carriers of (quantum) information, transmitting those carriers in a reliable way, and processing the information. Obviously, in order to maximize the communication rate, the duration of the carriers, naturally photons, should be as short as possible. However, while quantum communication channels such as optical fibres or free-space channels allow transmission of broadband, sub nanosecond photons, it is often impossible to generate or process such carriers. Of particular concern are key elements for quantum repeaters \cite{Sangouard2009} such as certain entangled photon pair sources \cite{Duan2001,McKeever2004,Chaneliere2007}, or quantum memories \cite{Lvovsky2009}, whose bandwidth is often limited by material constraints.

In this article, we study temporal/bandwidth conversion as a quantum information preserving transformation for photonic information carriers. On the one hand, increasing the optical bandwidth in an efficient way, i.e. decreasing the duration of an information carrier, will allow increasing the transmission rate over a single quantum channel when time-multiplexing several small bandwidth photon sources or photon pair sources. On the other hand, decreasing the bandwidth will allow mapping of large-bandwidth photonic carriers into quantum memories with limited spectral width. In view of this transformation, we investigate a photon-echo type quantum memory approach based on controlled reversible inhomogeneous broadening (CRIB)  of a large ensemble of atomic absorbers \cite{Moiseev2001,Nilsson2005,Kraus2006,Alexander2006,Tittel2009}.  Relaxing the requirement of symmetric inversion of atomic detunings $\Delta\rightarrow -\Delta$ during absorption of the light and recall, respectively, in a more general version with non-symmetric inversion $\Delta\rightarrow -\eta\Delta$ with compression factor $\eta\neq 1$, we find accelerated/decelerated rephasing of atomic coherences and thus temporal compression/decompression of the reemitted light field.  We analytically analyze the proposed scheme in optically thick atomic media with transverse and longitudinal inhomogeneous broadenings for \textit{recall efficiency}, \textit{fidelity}, and \textit{gain} of the transmission rate over a single channel through multiplexing. In transverse inhomogeneously broadened media, the atomic resonance frequencies vary normal (transverse) to the spatial coordinate z measured along the propagation direction of the light, and the absorption profile is independent of z. In longitudinal inhomogeneously broadened media, the resonance frequencies depend linearly on z. Assuming large optical depth for storage and retrieval, we find, for the case of transverse broadening, that the recall efficiency  is limited by the compression factor $\eta$, while it reaches unity in the case of a longitudinal broadened medium. We also find, for transverse broadening, that the fidelity of a recalled photonic time-bin qubit with the original qubit is one, regardless the compression factor, but that it is limited in the case of longitudinal broadening.
We point out that optical pulse compression has previously been considered using traditional photon echos and chirped excitation pulses \cite{Bai1984,Bai1986,Wang2000}. However, similar to data storage \cite{Ruggiero2009,Sangouard2010}, this approach is not suitable for temporal compression of quantum data. Quantum compression using CRIB has first been discussed in \cite{conference}, and first observations as well as numerical studies for the case of longitudinal broadening have recently been reported  \cite{Hosseini2009,Buchler2010}.

This article is organized in the following way. We will first discuss the figures of merit chosen to assess the performance of quantum compression. We will then introduce photonic \textsl{time-bin qubits} and present two ways to describe this basic unit of quantum information. This part is followed by a description of the standard CRIB-based quantum memory protocol, which employs a hidden symmetry in the equations describing the atom-light interaction during storage and recall. In order to assess the change in the recalled photonic quantum state for CRIB-based quantum compression, which does not involve time-reversed quantum dynamics, we have to solve the equations of motion. This is done in the next section, where we also derive the efficiency, fidelity and gain of quantum compression for the examples of transverse and longitudinal broadening. This is the main part of this article. It is followed by a brief feasibility study of our protocol in rare-earth-ion doped crystals. The article terminates with a conclusion.

\section{Figures of merit}
Temporal compression of quantum data is of importance to quantum communication, similar to temporal compression of classical data and classical (tele) communication. However, the criteria imposed on a "good" compression procedure are much more severe in the quantum case: reduced efficiency impacts irreversibly on the quantum information rate through photon loss, in opposition to amplitude loss in the classical case, which can be compensated by means of optical amplifiers (note that amplification is unsuitable for quantum communication, as stated in the no-cloning theorem \cite{Wootters1982,Dieks1982}). Furthermore, unpredictable modification of the input photonic quantum state $\ket{\psi_{in}}_{p}$ during compression results in an increased quantum bit error rate (QBER), or requires compensation via not-yet-practical quantum error correction \cite{Shor1995,Chuang1995}. In opposition, classical information, due to its digital nature, is much more tolerant to noise. In this article we therefore use the efficiency $\epsilon$, the fidelity $F$ (which specifies the unpredictable change of an input quantum state), and the gain $G$ (which derives from the efficiency and the compression parameter), as figures of merit to analyze the performance of quantum compression.

In the following, we assume pure photonic qubit input states. The compression operation yields a (possibly mixed) photonic output state characterized by a generally not normalized density matrix $\hat{\rho}_{out,p}$.
We define the efficiency $\epsilon$ of the compression  as
\beq\label{efficiency_definition}
\epsilon=tr(\hat{\rho}_{out,p})
\eeq
\noindent
The index "p" denotes "photon" and will be added henceforth to avoid confusion of photonic with atomic states. Furthermore, we use the following definition of the fidelity $F$ :
\beq\label{fidelity_definition}
F=_{p}\bra{\psi_{in}}\rho'_{out,p}\ket{\psi_{in}}_{p}
 \eeq
 \noindent where $\rho'_{out,p}$ is the renormalized and unitarily transformed density matrix
 \beq\label{rho out prime}
 \rho '_{out,p}=\frac{1}{\epsilon}U\rho_{out,p}U^+,
 \eeq
 \noindent
and $\rho_{out,p}$ describes the recalled and compressed photonic qubit. It is obtained from the total density matrix by tracing over the degrees of freedom related to the atomic system and possibly non-compressed photonic modes. This procedure is justified as one can experimentally restrict photon detection to the desired cases.

	Note that we allowed for deterministic unitary operations $U$ composed of rotations around $\sigma_x$, $\sigma_y$, or $\sigma_z$ in the qubit Hilbert space to maximize the fidelity. This is similar to quantum teleportation \cite{Bennett1993}, where, depending on the result of the Bell state measurement, bit flip, phase flip, or bit and phase flip operations have to be applied to the teleported state to recover the initial state.

Finally, in order to characterize the usefulness of quantum compression in view of enhanced qubit transmission rate, we define the gain as
\beq\label{gain_definition}
G=\epsilon\eta.
\eeq
\noindent
This reflects that the detection rate of qubits increases linearly with both the compression efficiency $\epsilon$ as well as the factor $\eta$ by which a qubit can be compressed, i.e. the number of qubits per unit time.

\section{Photonic time-bin qubits}

We are concerned with encoding of quantum information into qubits, i.e. quantum states that are described by
\beq\label{qubit}
\ket{\psi}=\alpha\ket{0}+ e^{i\phi}\beta\ket{1}.
\eeq
\noindent
The coefficients $\alpha$, $\beta$ and $\phi$ are real, $\alpha^2+\beta^2=1$, and the kets $\ket{0}$ and $\ket{1}$ form an orthogonal basis in a two-dimensional Hilbert space ($\braket{i}{j}=\delta_{ij}, i,j=[0,1]$).

\begin{figure}

  \includegraphics[width=0.35\textwidth]{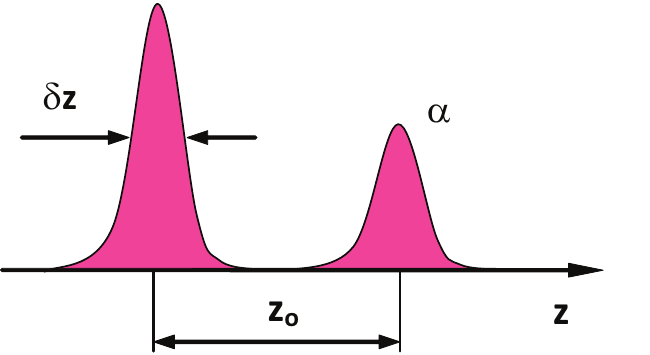}
  \caption{Schematical representation of a photonic time-bin qubit $\ket{\psi}=\alpha\ket{0}+\beta e^{i\phi}\ket{1}$, where $\alpha$ and $\beta $ are the amplitudes of photon wave packets propagating along the +z axis with relative phase $\phi$; $\delta z =c \delta t$, and $z_o=c \tau_o$.  $ \delta t$ and $\tau_o$ denote the temporal duration of the wave packets and the relative time delay, respectively.  }\label{figure_time-bin_qubit}
\end{figure}

In this article, we are specifically interested in so-called \textit{time-bin }qubits, where
the basis states $\ket{0}$ and $\ket{1}$ describe photon wavepackets localized at a particular position $z$ at early and late times $t=z/c$ and
$t'=z/c+\tau_{o}$, respectively. Differently stated, at a given time $t$, the photon is in a superposition of being at positions $z$ and $z'=z-z_o$ with $z_0=c\tau_o$, see Fig. \ref{figure_time-bin_qubit}. We will use these two pictures interchangeably in our analysis. Time-bin qubits have been shown to be very well suited for quantum communication over telecommunication fibres \cite{Tittel2001,Riedmatten2004}.
In the following, we will derive a description of time-bin qubits using a physical representation of the abstract qubit Hilbert space. We will limit ourselves to one polarization mode.

A general photonic wavepacket is described in \textit{momentum space} as \cite{Scully1997,Loudon2000}
\begin{equation}
\label{general photonic state}
\ket{\psi (t)}_p = \int_{ -\infty }^\infty{dk \tilde{g}
(k,t)} \hat {a}_{k}^ + \ket{\emptyset},
\end{equation}

\noindent
where $\hat{a}_{k}^ +$ ($\hat{a}_{k}$) denotes the creation (annihilation) operator for a photon with wave vector $k$, $[\hat{a}_{k},\hat{a}_{k'}^+ ] = \delta (k - k')$, and $\ket{\emptyset}$ is the vacuum state of light. For a more suitable description of our qubit states, we define operators creating forward and backward propagating photons at a particular position (we will index these two cases by \textit{f} and $+$, or \textit{b} and $-$, respectively):

\bea\label{position creation operators}
\hat{a}_+^ + (z) =\hat {a}_f^ + (z) = \sqrt {1 / 2\pi } \int_{0}^\infty
{dk} \exp \{ - ikz\}\hat {a}_{k}^+,\\
\hat{a}_-^ + (z) =\hat {a}_b^ + (z) = \sqrt {1 / 2\pi } \int_{-\infty}^{0}
{dk} \exp \{ - ikz\}\hat {a}_{k}^ +.
\eea
\noindent
The associated annihilation operators are defined by
$\hat {a}_{f}(z)= (\hat {a}_{f}^ + (z))^ + $ and $\hat {a}_{b}(z)= (\hat {a}_{b}^ + (z))^ + $, satisfying  the usual bosonic commutation relations. With these new operators, the basis states of a time-bin qubit can be expressed via the center $z=\sigma ct$ and $z-z_o=\sigma c(t-\tau_o)$ ($\sigma =[+,-]$) of the wavepackets, respectively, their extension $\delta z$ and carrier frequency $\omega_\sigma$:

\bwt
\bea
\label{temporal wavepacket1}
\ket{0}_{\sigma,p} &=& \int_{ - \infty }^\infty
{dz'} \exp\{-i\omega_\sigma (t -\sigma z'/c)\}g (ct - \sigma z',\delta z)\hat {a}_\sigma^ + (z')\ket{\emptyset},\\
&\equiv&\ket{t ,\delta z,\omega _ \sigma }_{\sigma,p}\nonumber
\eea
\bea
\label{temporal wavepacket2}
\ket{1}_{\sigma,p}&=& \int_{ - \infty }^\infty
{dz'} \exp\{-i\omega_\sigma (t -\tau_o - \sigma z'/c)\}g (ct -c\tau_o- \sigma z',\delta z)\hat {a}_\sigma^ + (z')\ket{\emptyset},\\
&\equiv& \ket{t -\tau_o,\delta z,\omega _ \sigma }_{\sigma,p} \nonumber
\eea
\ewt

\noindent
where $g (ct - z',\delta z)$ is the normalized envelope of the photon wave packet with length $\delta z$, which is related to its spectral function $\bar{g}(k,\delta k)$ by Fourier transformation: $g (z,\delta z)= \sqrt { 1 / 2\pi } \int_{ - \infty }^\infty
{dk \bar{g} (k,\delta k)\exp \{ - ikz} \}$. Note that the identification of the states $\ket{t,\delta z,\omega_\sigma}_{\sigma,p}$ and $\ket{t-\tau_o,\delta z,\omega_{\sigma}}_{\sigma,p}$ with $\ket{0}_\sigma$ and $\ket{1}_\sigma$, respectively, requires $\delta z \ll z_o=c\tau_0 $ so that the states are orthogonal and normalized.
Using this  notation, the general photonic wavepacket can then be expressed as
\bwt
\bea
\label{general photonic wavepacket}
\ket{\psi (t)}_p &=& \ket{\psi (t)}_{f,p}+\ket{\psi (t)}_{b,p}\nonumber\\
&=&\sqrt{1/2\pi}\sum_\sigma \int {dz} A_\sigma(t,z')\exp\{-i\omega_\sigma(t-\sigma z'/c)\}\hat {a}_{\sigma}^ +(z')\ket{\emptyset}.
\eea
\ewt
\noindent In the case of the initial, forward propagating time-bin qubit, the amplitude $A_+$ takes on the form

\bea
\label{initial envelope}
A_ + (t\to -\infty,z) &=&\sqrt {2\pi} \big \{\alpha g (ct-z,\delta z) + \exp \{i\omega _ + \tau
_o + i\phi \}\beta g(ct-z - c\tau _o ,\delta z )\big \}
\eea
\noindent and $
A_-(t\to -\infty,z )=0$, hence

\bea
\label{photonic qubit in time domain}
\ket{\psi_{in}(t)}_p &=&\alpha \ket{t,\delta z,\omega _ + }_{f,p} +
 e^{i\phi }\beta\ket{t -\tau_o ,\delta z,\omega _ + }_{f,p}.
\eea
\noindent
A schematical representation of a time-bin qubit assuming Gaussian shapes of the basis wave packets is given in Fig. \ref{figure_time-bin_qubit}.

\vspace{0.5cm}

\section{Quantum state storage based on symmetric CRIB}
\label{section symmetric CRIB}

An extensively investigated approach to quantum state storage is based on controlled reversible inhomogeneous broadening (CRIB) of an isolated absorption line \cite{Moiseev2001,Kraus2006,Alexander2006,Nilsson2005,Tittel2009}.  It has recently led to storage efficiencies up to 69\% \cite{Hedges2010}, and the possibility for random access quantum memory has been demonstrated \cite{Hosseini2009}. In the ideal,  standard CRIB scheme, the optical input data is launched along the forward (+z) direction into a resonant, optically thick atomic medium ($\alpha L\rightarrow\infty$, where $\alpha$ is the resonant absorption coefficient and $L$ is the length of the atomic medium). The spectral
width $\delta \omega_{in}$ of the input light should be narrow in comparison with the inhomogeneous broadening $\Delta _{inh}$ of the resonant atomic transition $1\leftrightarrow 2$, i.e. $\delta \omega _{in} < \Delta _{inh}$, and larger than the homogeneous line width $\gamma_{eg}$ of the optical transition, i.e. $ \delta \omega _{in} > \gamma_{eg}$.

All atoms are initially prepared in the pure state
$\ket{g} = \prod_{j = 1}^N {\otimes \ket{g}_j } $ (where $\ket{g}_j$  is a long-lived (ground) state of the j-th atom). Thus the initial light-atom state, denoted using the capital letter $\Psi$, (for $t\rightarrow -\infty$) is

\begin{equation}
\label{initial light atom state}
\ket{\Psi (t \to - \infty )}_f = \ket{\psi _{in} (t)}_{f,p}\otimes \ket{g}.
\end{equation}

After absorption of the light, the atomic coherences acquire phase factors $\exp \{ - i(\Delta _j + \omega _{eg} )t\}$, where $\Delta _j$ is the detuning of atom $j$ as compared to the central absorption frequency $\omega _{eg}$, leading to fast dephasing of the excited collective atomic coherence. The absorption process can be time reversed by inverting the atomic detunings at time $t_1$ (i.e. $\Delta_j\rightarrow -\Delta_j$) and applying a position dependent phase shift of $2kz$. This leads to rephasing of the collective atomic coherence, hence irradiation of an echo signal at time $t_{echo}$ in backwards direction as a perfect time reversed copy of the input data \cite{Moiseev2001}. This reversibility is based on a hidden symmetry of the equations describing the evolution of the slowly varying light and atomic parameters during storage and retrieval \cite{Kraus2006}.

Analyzing the evolution of the complete wavefunction (including the fast, time varying part $exp\{-ick_{eg}t\}$ where $ck_{eg}$ is the central frequency of the atomic transition), it was shown \cite{Moiseev2007} that the photon operators (in momentum space) of the input (\textit{forward}) and output (\textit{backward}) light fields are related by
$\exp \{ - ic(k + k_{eg} )t\}\hat {a}_{k + k_{eg} }^ + \to - \exp \{ic(k - k_{eg} )(t
- t_{echo} )\}\hat {a}_{k - k_{eg} }^ +.$
As we will show in more detail in Appendix A, this results for input photonic time-bin qubit states (Eq. \ref{photonic qubit in time domain})  to emerge as

\bwt
\bea
\label{photonic output qubit state in time domain}
\ket{\psi_{out}(t-t_{echo})}_{b,p} &=& -\bigg (\alpha \ket{t,\delta z,\omega _ - }_{b,p} +
e^{i(\phi+2\omega_{eg}\tau_o )}\beta\ket{t +\tau_o ,\delta z,\omega _ - }_{b,p}\bigg ),
\eea
\ewt
\noindent
where we have ignored a global phase shift, and where $\omega_-=2\omega_{eg}-\omega_+$ denotes the new carrier frequency. Hence, the recalled qubit state is associated with an exchange of the leading and trailing wavepackets, and the now leading wavepacket (with probability amplitude proportional to $\beta$) has acquired an additional phase that depends on the time delay $\tau_o$ between the wavepackets, and the atomic transition frequency $\omega_{eg}$ (see also \cite{Gisin2007}).  Please note that not only the order of the wavepackets change, but that each wavepacket's temporal envelope is also time-reversed. For simplicity of notation, we will henceforth restrict our investigation to time-symmetric, Gaussian shaped basis wavepackets described by $g (ct-z,\delta z)=\sqrt[4]{2/(\pi(\delta z)^2)}
\exp \{ - \left({c\tau / \delta z} \right)^2\}$, where $\tau=t-z/c$.

Returning to the abstract qubit notation, we find that the initial state
$\ket{\psi_{in}}=\alpha\ket{0}+ e^{i\phi}\beta\ket{1}$ is transformed in the quantum memory
into $\ket{\psi_{out}}=\alpha\ket{1}+e^{i(\phi+2\omega_{eg}\tau_o )}\beta\ket{0}$,
i.e. that the two states are related by a deterministic unitary transformation $T$
\beq
\label{output wave function in symmetric CRIB}
\ket{\psi_{out}}=T\ket{\psi_{in}}=e^{i\omega_{eg}\tau_o}\sigma_xR_z(\theta)\ket{\psi_{in}},
\eeq

\noindent
 where $R_z(\theta )=e^{i\textstyle{\theta\over 2}\sigma_z}$ denotes
a rotation of $\theta =2\omega_{eg}\tau_o$ around $\sigma_z$,
and
$\sigma_x$
 is the bit flip operator.

Hence, for the ideal standard CRIB protocol,
we find the efficiency to be one. Furthermore, from Eq. (\ref{rho out prime}),
we find $U=T^{-1}$, hence $\rho_{out,p}'=\rho_{in,p}$ and thus F=1. Please note that two subsequent storage sequences lead to compensation of the bit flip as well as the additional phase factor of $2\omega_{eg}\tau_o$.

Finally, being obvious in the case of "no compression" and unity efficiency, we note that the gain $G$ is one.

\section{Quantum compression based on generalized CRIB}

In the CRIB protocol described above, the efficiency, fidelity, and gain can be derived using arguments stemming from symmetries in the equations of motion. Relaxing the perfect reversibility of atomic detunings by introducing a more general relation between the initial ($t<t_1$) and output ($t>t_1$) spectral detunings
\beq\label{Delta to -eta Delta}
\Delta_j(t>t_1)= -\eta\Delta_j(t<t_1),
\eeq
\noindent
 with $\eta$ being the compression parameter, this symmetry-based approach is no longer possible. Here, we pursue the following approach: We start with a general photonic qubit state  (Eq. \ref{qubit}) described in the abstract two-dimensional Hilbert space. We then express the qubit state using a physical description in position space (Eqs. (\ref{temporal wavepacket1}, \ref{temporal wavepacket2}), and (\ref{photonic qubit in time domain}), respectively). Using the equations of motion, which we will introduce in the following section, we then calculate the state of the combined atom-photon system after quantum compression. From Eq. (\ref{photonic output qubit state in time domain}), we find the single photon density matrix. Redefining new basis states $\ket{0}$ and $\ket{1}$, determined by the compression parameter, we then express the output state in qubit state notation, which finally allows us to calculate our figures of merit (Eqs. (\ref{efficiency_definition}), (\ref{fidelity_definition}), and (\ref{gain_definition}), and the unitary operation $U$ (see Eq. (\ref{rho out prime})).

\subsection{Basic equations}

The interaction of the quantum fields with atoms is given by the Hamiltonian \cite{Scully1997,Loudon2000,Moiseev2007}

\begin{equation}
\label{total Hamiltonian}
\hat {H}(t) = \hat {H}_{0} + \hat {H}_{1}(t),
\end{equation}
\noindent
where
\bea\label{Hamiltonian 0}
\hat {H}_0 = \hbar\omega_{eg}\bigg ( \sum_{\sigma=\pm}\int_{-\infty}^{\infty}dz \widehat{a}^+_{\sigma}(z)\widehat{a}_{\sigma}(z) + \sum_{j=1}^N {P}_{ee}^j\bigg ),
\eea
and

\bea
\label{Hamiltonian1}
\label{interaction Hamiltonian}\hat {H}_1(t)&=&\nonumber\\
&-& \hbar g\sqrt{2\pi}\sum\limits_{j = 1}^N \bigg \{ \big \{\hat {a}_{f}(z_j)+\hat {a}_{b}(z_j) \big \}
\hat {P}_{eg}^j + h.c. \bigg \}\nonumber\\
\label{line broadening Hamiltonian}&+&\sum_{j = 1}^N {\big (\hbar \Delta _j (t) +\delta  E_{eg}^j (t)\big )\hat {P}_{ee}^j}\nonumber\\
\label{propagation Hamiltonian}
&-&i\hbar c \sum_{\sigma=\pm}\sigma\int_{-\infty}^{\infty}dz \hat{a}^+_{\sigma}(z)\frac{\partial}{\partial z}\hat{a}_{\sigma}(z).
\eea
\noindent
$\hat{H}_0$ describes the total number of excitations in the system and commutes with the total Hamiltonian $\hat{H}$. The first term in $\hat{H}_1(t)$ describes the atom-field interaction with atomic operator $\hat {P}_{mn}^j=\ket{m}_{jj}\bra{n}$. $N$ denotes the number of atoms, $g(\omega ) \cong g = id_{ge} (\textstyle{{\omega _{eg} } \over
{2\hbar \varepsilon _o S}})^{1 / 2}$, d$_{12}$ is the dipole moment of the
atomic transition $\left| g \right\rangle \leftrightarrow \left|e\right\rangle $,
S is the cross section of the light beams, $\varepsilon _o $
is the electric permittivity, and $\hbar $ is Planck's constant divided by 2$\pi $. The second term in Eq. \ref{line broadening Hamiltonian} describes inhomogeneous and homogeneous atomic line broadening within a unitary approach to quantum evolution. Indeed, it has been shown \cite{Klauder1962,Salikhov1976} that the decay rate of the atomic coherence $\gamma_{eg}$ can be calculated by a statistical average over local stochastic phase fluctuations
$\delta  \phi _{eg}^j (t
,t ') = \int_{t '}^t {dt "} \delta  E_{eg}^j(t) /\hbar$
of the atomic transition:

\bea
\label{homogeneous broadening}
\big\langle \exp \{ - i  \delta \phi _{eg}^j (t
,t ')\}\big\rangle = \exp \{ -
\gamma _{eg} (t - t ')\},
\eea

\noindent
where
$\big\langle \delta \hat\phi _{eg}^j (t
,t ')\big\rangle = 0$.
These fluctuations are due to  interaction between the absorbers and its environment. 
Finally, the third term of $\hat{H}_1(t)$ takes into account the spectrum of the localized photonic wave packets (here expressed through position operators.)

Using the total Hamiltonian  (\ref{total Hamiltonian}), the initial photon-atom state (\ref{initial light atom state}) will evolve into

\begin{equation}
\label{general state}
\left| {\Psi (t)} \right\rangle = \left| {\Psi (t)} \right\rangle _p +
\left| {\Psi (t)} \right\rangle _a ,
\end{equation}

\bea
\label{general photon state1}
\left| {\Psi (t)} \right\rangle _p = \ket{\psi (t)}_p\otimes\ket{g} ,
\eea
\noindent
(with $\ket{\psi (t)}_p$ as defined in Eq. (\ref{general photonic state})) and
\begin{equation}
\label{general atomic state}
\left| {\Psi (t)} \right\rangle _a = \sum\limits_{j = 1}^N  b_j (t)\hat
{P}_{eg}^j \ket{g}
\otimes \ket{\emptyset},
\end{equation}
\noindent
with $b_j(t)$ being the probability amplitude for atom $j$ to be in the excited state. Initially $b_j (t \to - \infty )
=0$,  i.e. all atoms are in the ground state, and $ A_ - (t, \to - \infty, z)) = 0$, i.e. all backward modes are empty.

In the Schr\"{o}dinger equation with the Hamiltonian introduced in (Eq. \ref{total Hamiltonian}), the quantum evolution during absorption of the forward propagating light field (${\sigma}=+$ for $t<t_1$) and for re-emission of the backward propagating light field  (${\sigma}=-$ for $t>t_1$) is given by

\bwt
\bea
\label{quantum evolution of field}
\big (\textstyle{\frac{1}{c} \frac{\partial }{\partial t}+{\sigma}
\frac{\partial}{\partial z}}\big )
A_ {\sigma} (t,z) &=& i(\pi n_o S
g^\ast / c)\exp \{i\omega _{\sigma} (t-{\sigma}z/c) \}b(t,z),\\
\label{quantum evolution of atoms}
\textstyle{\partial \over {\partial t }}b_j (t) &=& - i\{\omega _{eg} +
\Delta _j(t) + \delta \Delta _{eg}^j (t)\}b_j (t) +
igA_ {\sigma} (t ,z_j )\exp \{ - i\omega _ {\sigma} (t-{\sigma}z_j/c)\},
\eea
\ewt

\noindent
where  $b(t ,z) = (n_o S)^{ - 1}\sum\nolimits_{j = 1}^N
{b_j (t )\delta (z - z_j )} $ is a collective atomic variable describing the averaged, position dependent atomic coherence, and $n_o$ is the atomic density. Using these equations, we will now evaluate quantum compression of  photonic time-bin qubits in atomic systems with \textit{transverse} and \textit{longitudinal} inhomogeneous broadening.

\subsection{Transverse broadened media}

\begin{figure}
  \includegraphics[width=0.75\textwidth]{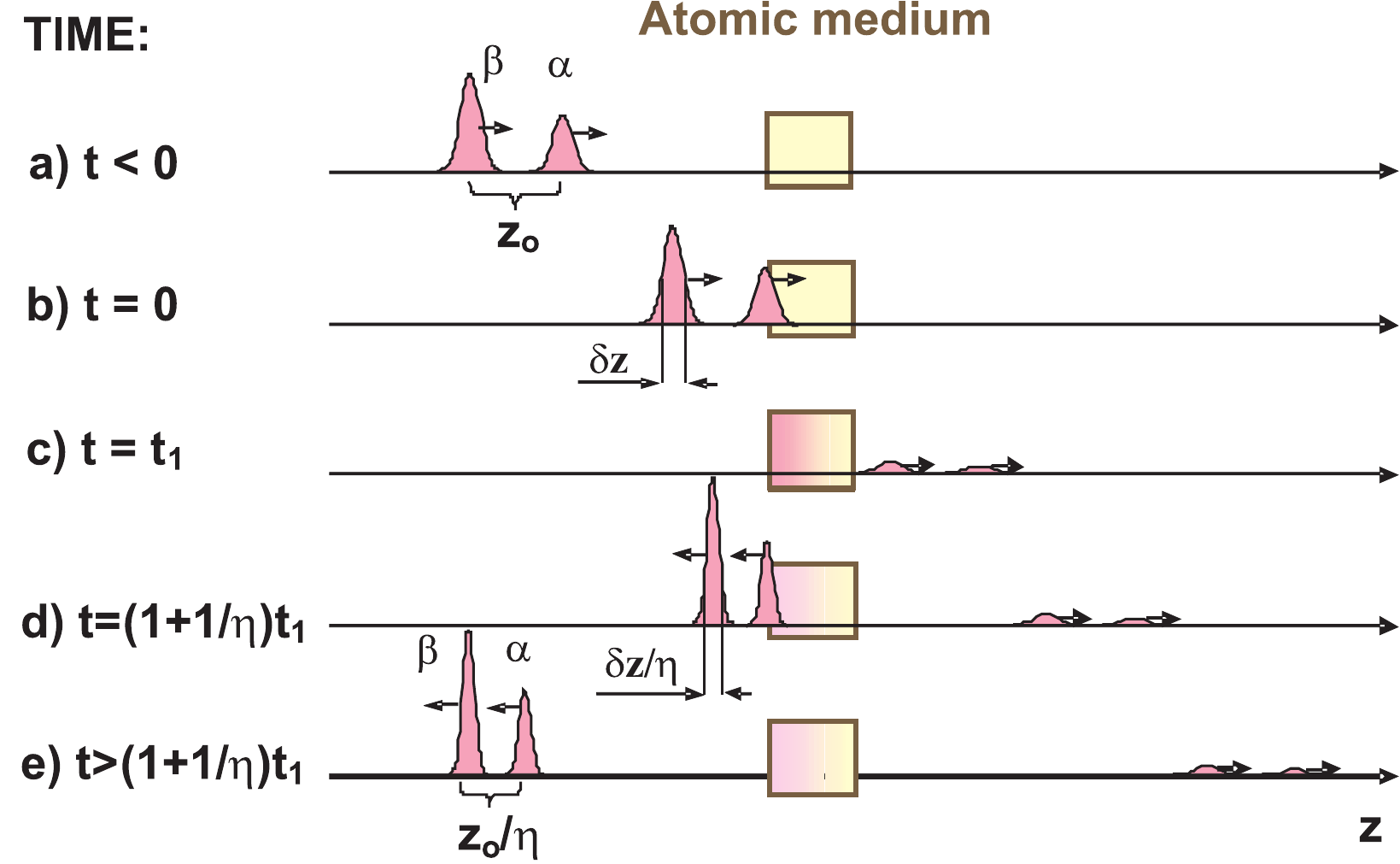}\\
  \caption{Schematics showing different instances in the quantum compression protocol in the case of transverse broadening. The figure depicts forwards and backwards propagating photonic wave packets, and atomic excitation.}
  \label{figure_setup_transverse_scheme}
\end{figure}

\textit{Transverse} broadening is the inhomogeneous broadening occuring naturally. In such media, the atomic absorption profile is independent of the spatial coordinate z measured along the propagation direction of the light. In rare-earth-ion doped inorganic crystals (RE crystals) with low site symmetry and sufficiently narrow absorption line, \textit{controlled transverse broadening} can be introduced through the Stark effect \cite{Macfarlane2007} by applying an electric field gradient \textit{transverse} to the propagation direction of light. For RE ions implemented in a glassy host, e.g. an optical fibre, controlled transverse broadening is already obtained for uniform electric fields, due to the random orientation and magnitude of the electric dipole moments of the RE ions \cite{Hastings-Simon2006}.  In the following, we assume a broadened medium with Lorentzian lineshape
$ G(\Delta /\Delta_{inh} )=\Delta_{inh} /((\Delta_{inh}^2+\Delta^2)\pi)$ with inhomogeneous spectral width $\Delta_{inh}\gg c\delta k$. The center of the first wave packet $\ket{0}_{f,p}$ enters the atomic medium at $t_0=0$ (see Fig. \ref{figure_setup_transverse_scheme}). During the interaction with the medium, the light field is partially or completely absorbed, depending on the optical depth. At time $t_{1}$, after sufficient dephasing, we  change the detuning $\Delta_j$ to $-\eta\Delta_j$, and we also apply a position dependent phase shift of $(2k_{eg}-\delta k)z$, which allows phase matching for the retrieval of the light field in backwards direction \cite{Moiseev2004,Moiseev2004b, Nilsson2005,Kraus2006}.
$\delta k$ describes a small deviation from the perfect phase matching. This leads to reemission of the light field, with the center of the (now trailing) wavepacket exiting the medium at time $t=t_{echo}$.  The atom-light state for $t\gg t_{echo}$ is then given by

\bea
\label{total output state}
\ket{\Psi(t\gg t_{echo})}&=&\ket{\psi_{f}(t)}_p\otimes\ket{g}
+ \ket{\Psi(t)}_a
+\ket{\psi_{b}(t)}_p\otimes\ket{g}.
\eea
\noindent
As visualized in Figs. \ref{figure_setup_transverse_scheme}c, d and e, the first term describes the damped, non-absorbed photonic component $\ket{\psi_{f}(t)}_p=exp\{-\alpha_oL/2\}\ket{\psi_{in}(t)}_p$ that continued to travel in forward direction behind the atomic medium, the second term the remaining atomic excitation, and the third term denotes the now backwards moving, retrieved photon field, which is the subject of our investigation. $\alpha_o$ is the on-resonant absorption coefficient, $L$ the length of the medium, and $\alpha_oL$ the optical depth. As shown in Appendix \ref{AppendixA}, the amplitude of the retrieved photon field, assuming for simplicity a Gaussian spectral shape, is described in abstract qubit notation by:
\bwt
\bea\label{final solution for Gaussian qubits transverse}
\ket{\psi _{b} (t)}^{(t)} _p
 = R^{(t)} \bigg \{
 e^{  -(1 +1/ \eta )\gamma _{eg} \tau_o }\alpha
\ket{t ,\delta z',\omega _ - }_b
+ e^{i\phi'}\beta\ket{t +\tau_o',\delta z',\omega _ - }_b \bigg \},
\eea
\ewt
\noindent
where the superscript "(t)" denotes transverse broadening, where we ignored a global phase shift, and where $\phi'=\phi+(1+1/\eta)\omega_{eg}\tau_o$. Furthermore,

\bwt
\beq
\label{temporal compressed wavepackets transverse}
\ket{t^{(m)},\delta z',\omega _ - }_b = \sqrt{\eta}\int_{ - \infty }^\infty
{dz'} \exp\{-i\omega_\sigma (t^{(m)} -t_{echo} + z'/c)\}g (ct^{(m)} -ct_{echo} +z',\delta z')\hat {a}_b^ + (z')\ket{\emptyset},
\eeq
\ewt
\noindent
where  $t^{(m)}$ denotes the early or late wave packet ($t^{(m)}\in [t, t +\tau_o']$),
 $t_{echo} = (1 + 1 / \eta ) t_1-\delta t_R$,
 $\delta t_R = \textstyle{1 \over 2}\gamma _{eg} [1 + \eta ](\delta z'/ c)^2$,
 $\tau_o'=\tau_o/\eta$ and $\delta z'=\delta z/\eta$. The factor
\bea
\label{inefficiencies}
R^{(t)} &=&\Gamma_{\gamma_{eg}}(t_1)M^{(t)}(\delta k,\alpha_oL)\epsilon_o^{1/2}(\eta)
\eea
\noindent
combines different sources that affect the efficiency of the quantum compression through phase relaxation $\gamma_{eg}$, phase mismatch $\delta k$, and limited optical depth $\alpha_oL$. Its upper limit, given by $\epsilon_o$ (not to be confused with the electric permittivity $\varepsilon_o$), is determined by the compression parameter $\eta$. Specifically, we find:

\beq
\label{Gamma}
\Gamma_{\gamma_{eg}}(t_1)=\exp \{
 - (1 + 1/\eta )\gamma _{eg}  (t_1-\tau_o
 -\textstyle{1 \over 2} \eta \delta t_R)\},
\eeq

\beq\label{M}
M^{(t)}(\delta k, \alpha _oL) = \frac{1 -
\exp \{ - \textstyle{1 \over 2}(1 + 1 / \eta )\alpha _o L + i\delta
kL\}}{1 - 2i(\delta k / \alpha _o )\eta / (\eta + 1)},
\eeq

\beq\label{epsilon_0}
\epsilon_o ^{1/2}(\eta)=\frac{2\sqrt \eta }{(\eta + 1)}.
\eeq

Let us now relate Eq. (\ref{final solution for Gaussian qubits transverse}) with our figures of merit. First, it is important to note that Eq. (\ref{final solution for Gaussian qubits transverse}) describes again a photonic time-bin qubit with modified separation and width of basis states. Indeed, comparing the new with the old basis states (Eq. (\ref{photonic qubit in time domain})), we find that the separation between the two basis wavepackets has changed from
$z_o$ to $z_o'=z_o/\eta$, and their extension from $\delta z$ to $\delta z'=\delta z/\eta$, while the wave packet amplitude has changed $\propto \eta^{1/2}$  as visualized in Figs. \ref{figure_setup_transverse_scheme} d and e.

Second, it is important to realize that the quantum information encoded into a qubit is independent of the physical realization of its abstract basis states $\ket{0}$ and $\ket{1}$. This allows us to relabel the basis states after compression, i.e. $\ket{t,\delta z',\omega_-}\rightarrow\ket{1}$, and $\ket{t+\tau_o',\delta z',\omega_-}\rightarrow\ket{0}$, resulting in:

\bea\label{compressed qubit in Hilbert space notation}
\ket{\psi}_p
 = R^{(t)}\bigg (
 e^{  -(1 +1/ \eta )\gamma _{eg} \tau_o }\alpha
\ket{1}
+ e^{i\phi'}\beta\ket{0}  \bigg ).
\eea
\noindent
Now, taking the trace of $\rho_{out,p}=\ket{\psi}_{pp}\bra{\psi}$ we find the efficiency
according to Eq. (\ref{efficiency_definition})
\begin{equation}
\label{recall efficiency}
\epsilon^{(t)}=
(\Gamma_{\gamma_{eg}}(t_1))^2 \vert M^{(t)}(\delta k,\alpha_oL)\vert ^2 \epsilon_o(\eta) \{\alpha ^2
 e^{ -2 (1 +1/ \eta )\gamma _{eg} \tau_o }
+ \beta ^2 \}.
\end{equation}
\noindent

We now discuss our figures of merit for some particular cases. Ignoring phase relaxation (i.e. $\gamma_{eg}=0$) and assuming perfect phase matching (i.e. $\delta k=0$) we find the efficiency to be given by:

\begin{equation}
\label{transverse ideal efficiency}
\epsilon^{(t)}=
\frac{4\eta }{(\eta + 1)^2}\big (1 - \exp \{ - \textstyle{1 \over 2}(1 + 1 / \eta
)\alpha _o L\}\big )^2.
\end{equation}

\begin{figure}
  \includegraphics[width=0.45\textwidth]{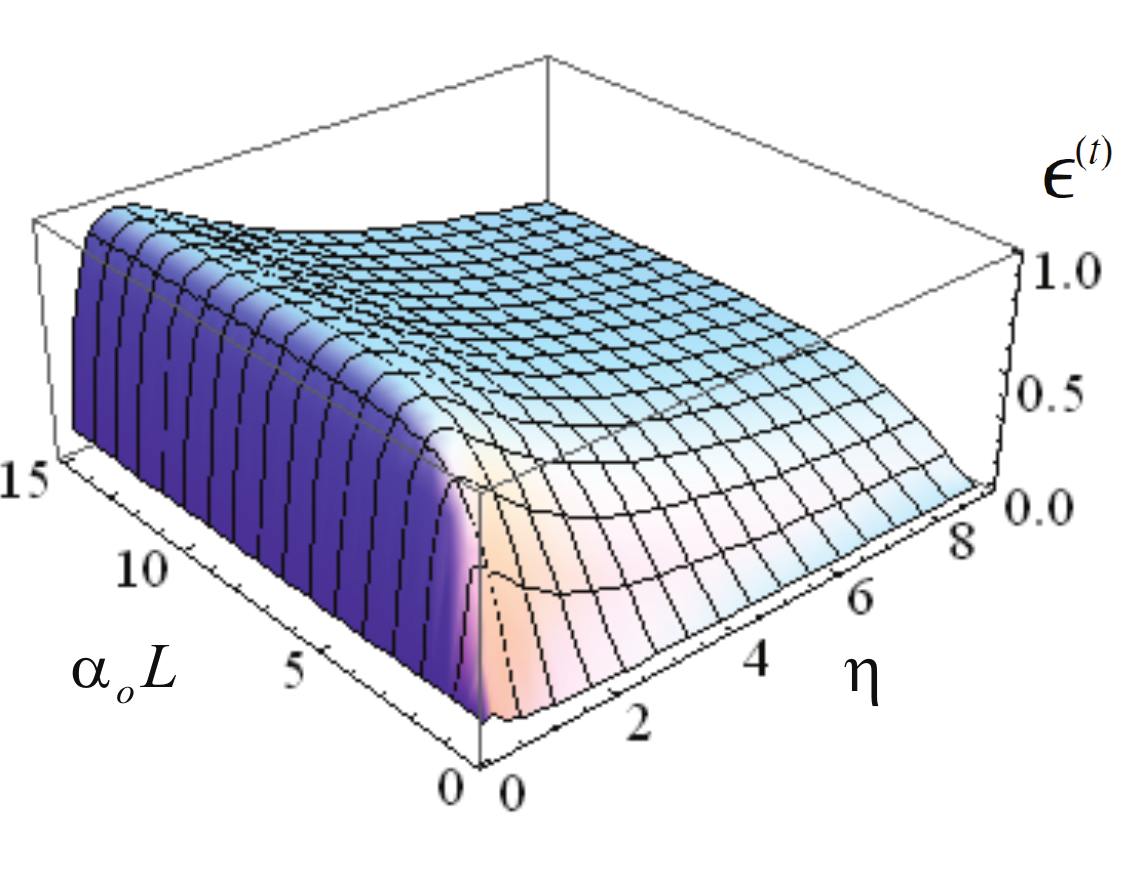} 
  \caption{Recall efficiency for transverse broadening as a function of the compression parameter $\eta$ and the optical depth $\alpha_o L$ (see Eq. (\ref{transverse ideal efficiency})).}
  \label{figure_efficiency_transverse_scheme}
\end{figure}

\noindent
 It is depicted in Figure \ref{figure_efficiency_transverse_scheme} as a function of the optical depth $\alpha_0L$ and the compression parameter $\eta$. First, we note that for standard CRIB (without compression, i.e. $\eta=1$), we find the previously published result $\epsilon^{(t)} = (1-\exp\{-\alpha_oL\})^2$ \cite{Moiseev2004,Sangouard2007}. Second, we point out that for a given compression factor, the maximum efficiency is always obtained for infinite optical depth, and hence the upper limit $\epsilon_{max}$ is only determined by the compression factor:
\beq
\label{efficiency fundamental limit transverse broadening}
\epsilon_{max}^{(t)}(\eta)=\frac{4\eta}{(1+\eta)^2}.
\eeq

\noindent To give an example, a quantum compression with $3-2\sqrt{2}\leq\eta\leq 3+2\sqrt{2}$ results in a recall efficiency limited to 50\%.

To assess the fidelity, we compare Eq. (\ref{compressed qubit in Hilbert space notation}) with the initial qubit state in Eq. (\ref{qubit}). We find that the re-normalized, compressed qubit state $\ket{\psi}_p$ can be unitarily transformed into the initial state by a $\sigma_z$ rotation of angle $\phi'-\phi=(1+1/\eta)\omega_{eg}\tau_o$ and a bit flip operation $\sigma_x$. Furthermore, one should take into account an additional $\sigma_x$ rotation to compensate for the amplitude reduction factor $\exp\{-(1+1/\eta)\gamma_{eg}\tau_o\}$ arising in the case of significant phase relaxation during time $(1+1/\eta)\tau_o$. This immediately implies that the fidelity as defined in Eq. (\ref{fidelity_definition}) is always one, regardless the compression factor, phase relaxation, or phase mismatch. Note that the fact that atomic phase relaxation does not affect the fidelity in time qubit storage has been inferred from stimulated photon echo experiments with intense light pulses \cite{Staudt2007}.

\begin{figure}
  \includegraphics[width=0.45\textwidth]{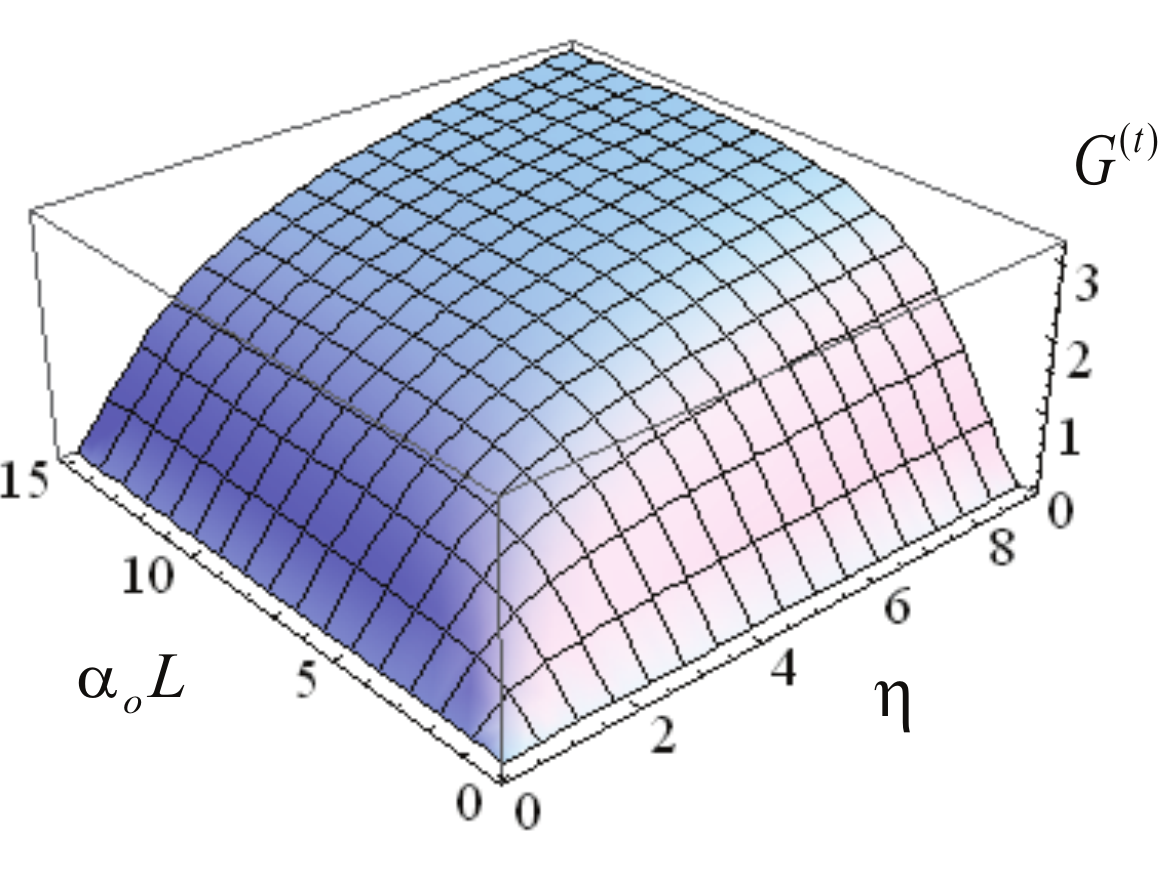} 
  \caption{Gain for transverse broadening as a function of the compression parameter $\eta$ and the optical depth $\alpha_o L$.}
  \label{figure_gain_transverse}
\end{figure}

Figure \ref{figure_gain_transverse} depicts the gain as a function of optical depth and compression parameter for transverse broadened media. As an example, assuming $\alpha_o L$=2, we find an increased communication rate, i.e. $G^{(t)}>$1, for $\eta\gtrsim$ 1.7. For large (infinite) optical depth (i.e. maximum efficiency as described by Eq. (\ref{efficiency fundamental limit transverse broadening})), the gain is upper bounded by
\beq
G^{(t)}(\alpha L\rightarrow\infty)=4\eta^2/(1+\eta)^2|_{\eta\gg 1}=4,
\label{max_gain_t}
\eeq
\noindent
a modest, yet significant improvement over quantum communication schemes without compression.

To finish this section, let us briefly inspect Eqs. (\ref{inefficiencies})-(\ref{epsilon_0}) in view of symmetry between compression and decompression, i.e. under exchange of $\eta$ to $1/\eta$. First, we recall that the fidelity is one, regardless the compression factor: $F(\eta )=F(1/\eta )=1$. Second, we find that the upper limit of the efficiency is symmetric with respect to compression/decompression: $\epsilon_{max}^{(t)}(\eta )=\epsilon_{max}^{(t)}(1/\eta )< 1$. However, limited optical depth, phase mismatch, and atomic phase relaxation break the symmetry in the efficiency. It should be possible to demonstrate this surprising behavior using only symmetry arguments and the equations of motion describing the recall efficiency, without having to solve the equations. This is likely to lead to a more profound understanding of the physical principles of the here studied compression/decompression protocol.

\subsection{Longitudinally broadened media}

\begin{figure}
  \includegraphics[width=0.75\textwidth]{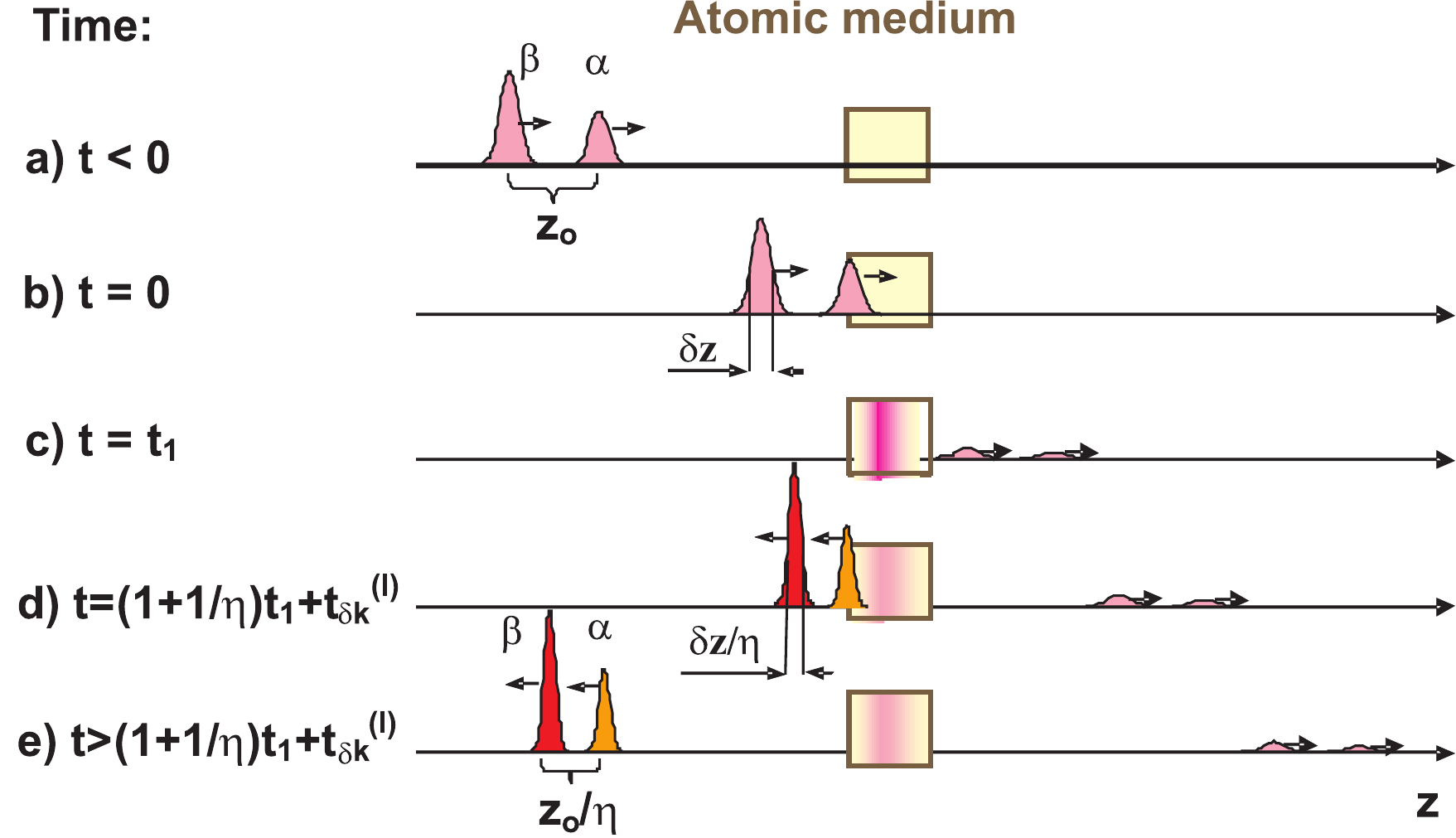}\\
  \caption{Schematic showing different instances in the quantum compression protocol in the case of longitudinal broadening. The figure depicts forwards and backwards propagating photonic wave packets, and atomic excitation. Note the difference of the re-emission time $t_{echo}$ and the different localization of atomic excitation compared to the transverse case (see Fig. \ref{figure_setup_transverse_scheme}). The change of colour of the irradiated wave packets denotes additional frequency shifts (see Eq.(\ref{delta_omega12})).}
  \label{figure_setup_longitudinal_scheme}
\end{figure}

\textit{Longitudinal} broadening is an inhomogeneous broadening that can not be found naturally. In this case, for each position z in the medium, the atomic absorption
profile is given by a narrow line (here for simplicity assumed to be homogeneously broadenend), whose center frequency varies monotonously with z. Longitudinal broadening can be realized through the Stark effect in non centro-symmetric RE ion doped inorganic crystals by applying a electric field gradient \textit{longitudinal} to the propagation direction of light \cite{Alexander2006}. In the following, we assume absorption lines whose detuning with respect to the light carrier frequency varies linearly with position $z$ within the crystal, which extends from $z=-L/2$ to $z= +L/2$:
\beq
\label{longitudinal broadening}
\Delta =-\chi z.
\eeq
\noindent
The initial inhomogeneous absorption line, obtained after integration over all atomic positions $z$, is assumed to be broad compared to the spectrum of the photonic wavepacket: $\Delta_{inh}=\chi L\gg c\delta k$. A schematics of the compression procedure depicting relevant instances of the protocol is given in Fig. \ref{figure_setup_longitudinal_scheme}.

A similar approach as in the case of transverse broadening leads to the following wavefunction for the backwards emitted light field (for more details see Appendix \ref{AppendixB}):

\bwt
\bea\label{final solution for Gaussian qubits longitudinal}
\ket{\psi _{b} (t)}^{(l)} _p
 = R^{(l)} \bigg \{
 e^{  -(1 +1/ \eta )\gamma _{eg} \tau_o }\alpha
\ket{t,t-t_{\delta k}^{(l)},\delta z',\omega_-}_b
+ e^{i\phi '}\beta\ket{t,t-t_{\delta k}^{(l)} +\tau_o',\delta z',\omega_-}_b \bigg \},
\eea
\ewt
\noindent

\noindent
where the subscript "(l)" denotes longitudinal broadening and where we ignored again a global phase shift. Furthermore,

\bwt
\bea
\label{temporal compressed wavepackets longitudinal}
\ket{t,t^{(m)},\delta z',\omega _ -}_b &=&\nonumber\\
\sqrt{\eta}\int_{ - \infty }^\infty
{dz'} \exp\{i\Phi(t+z'/c,\eta)\}\exp\{-i\omega_- (t^{(m)} -t_{echo}+ z'/c)\}g (ct^{(m)} -ct_{echo} +z',\delta z')\hat {a}_b^ + (z')\ket{\emptyset},
\eea
\ewt
\noindent
and $t^{(m)}\in [t-t_{\delta k}^{(l)} +\tau_o', t-t_{\delta k}^{(l)}]$
with

\beq
t_{\delta k}^{(l)} = \textstyle{1 \over \eta \Delta_{inh}}(\delta kL + 2(\zeta /
\chi )(1 - 1 / \eta )).
\eeq
\noindent
The parameters $\delta z'=\delta z/\eta$, $\tau_o'=\tau_o/\eta$, and $\phi '$ are the same as in the transverse case, and $\zeta= \pi n_o Sg^2 / c$. Furthermore,

\bea
\label{inefficiencies longitudinal}
R^{(l)} &=&\Gamma_{\gamma_{eg}}(t_1+t_{\delta k}^{(l)})M^{(l)}(\zeta/\chi,\eta)
\eea
\noindent
with

\bea
\label{M(l)}
M^{(l)}(\zeta/\chi,\eta)
&=& \sqrt{(1-\exp\{-\textstyle{2\pi\zeta \over \eta\chi}\})(1-\exp\{-\textstyle{2\pi\zeta \over \chi}\})}.
\eea
\noindent
Finally, the time dependent phase $\Phi(\tau=t+z/c,\eta)$ is given by

\bea
\label{varphi(tau)}
\Phi (\tau,\eta)=
\frac{\zeta}{\chi}ln \bigg \{\frac{\left( {\textstyle{1 \over 2}\eta\Delta_{inh} \abs { \tau - t_1 +
\tau _{m}' - \frac{\delta k L}{\eta \Delta_{inh} }} } \right)}{\left( {\textstyle{1 \over 2}\eta\Delta_{inh} (\tau - t_1 + \tau _{m}' )}\right)^{1/\eta}}\bigg \},
\eea

\noindent
where $\tau_{m}' =  \tau_m /\eta^2$, and $\tau_m =2 \frac{\zeta}{\chi\Delta_{inh}}$. For time-bin qubits with sufficiently narrow basis wavepackets, this nonlinear phase change translates into a different frequency change of each wavepacket ($\delta\omega_0$, $\delta\omega_1$) as compared to the carrier frequency. From $\delta\omega (t)=d\Phi(t)/dt$ we find:

\bea
\label{delta_omega12}
\delta \omega _0 &=& \frac{\zeta}{\chi}\bigg [\frac{1}{t_1 - \tau _o + \tau _m
 + \delta k / \chi} - \frac{\eta }{t_1 - \tau _o + \tau _m
}\bigg ],\nonumber\\
\delta \omega _1 &=& \frac{\zeta}{\chi}\bigg [\frac{1}{t_1 + \tau _m  + \delta
k / \chi} - \frac{\eta }{t_1 + \tau _m }\bigg ].
\eea
\noindent

\begin{figure}
  \includegraphics[width=0.45\textwidth]{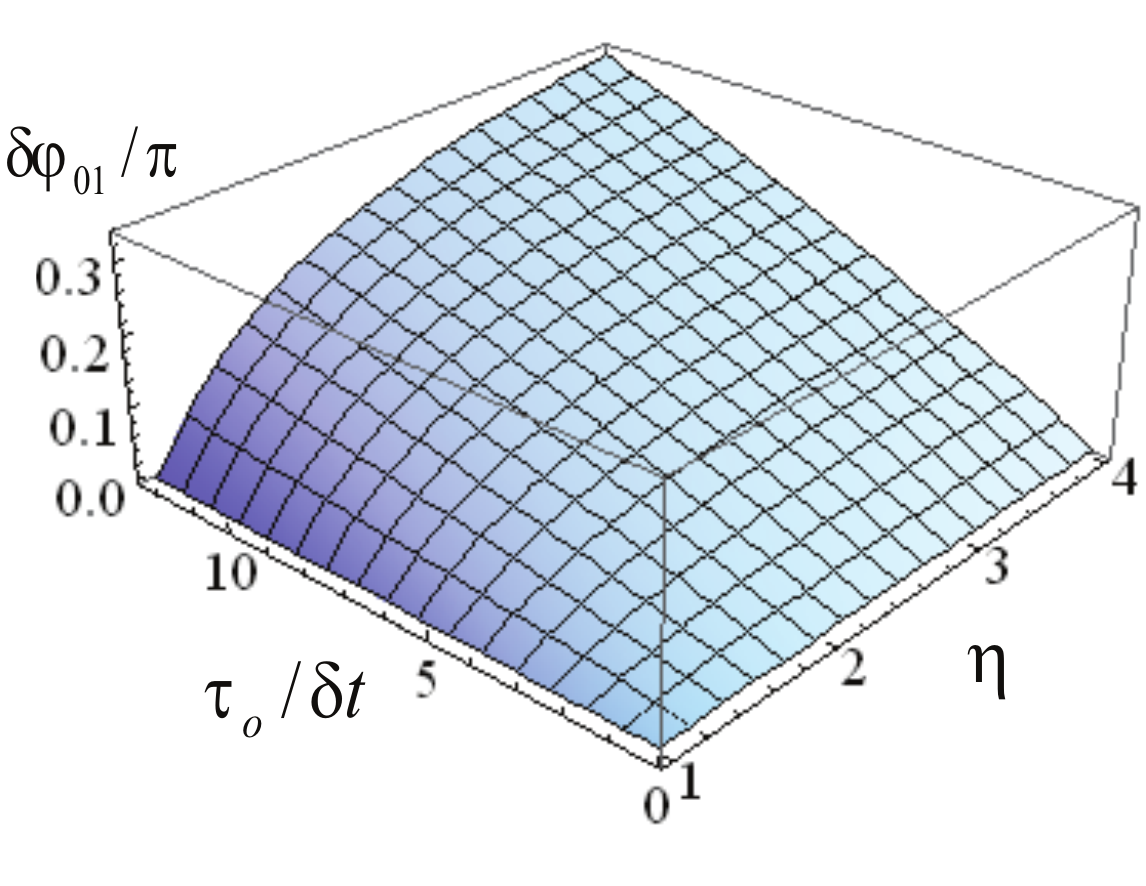}\\
  \caption{Phase shift $\delta \varphi_{01}$ for quantum compression as a function of the time delay $\tau_o$ and the compression parameter $\eta$, with
  dephasing time $t_1=20\delta t$, $\Delta_{inh}\delta t=10$, and $2 \pi \zeta/\chi=6\pi$. $\delta t$ denotes the temporal duration of the initial basis wavepackets, and $2 \pi \zeta/\chi$ is the effective optical depth.}
  \label{total_phase_shift1comp}
\end{figure}

\begin{figure}
  \includegraphics[width=0.45\textwidth]{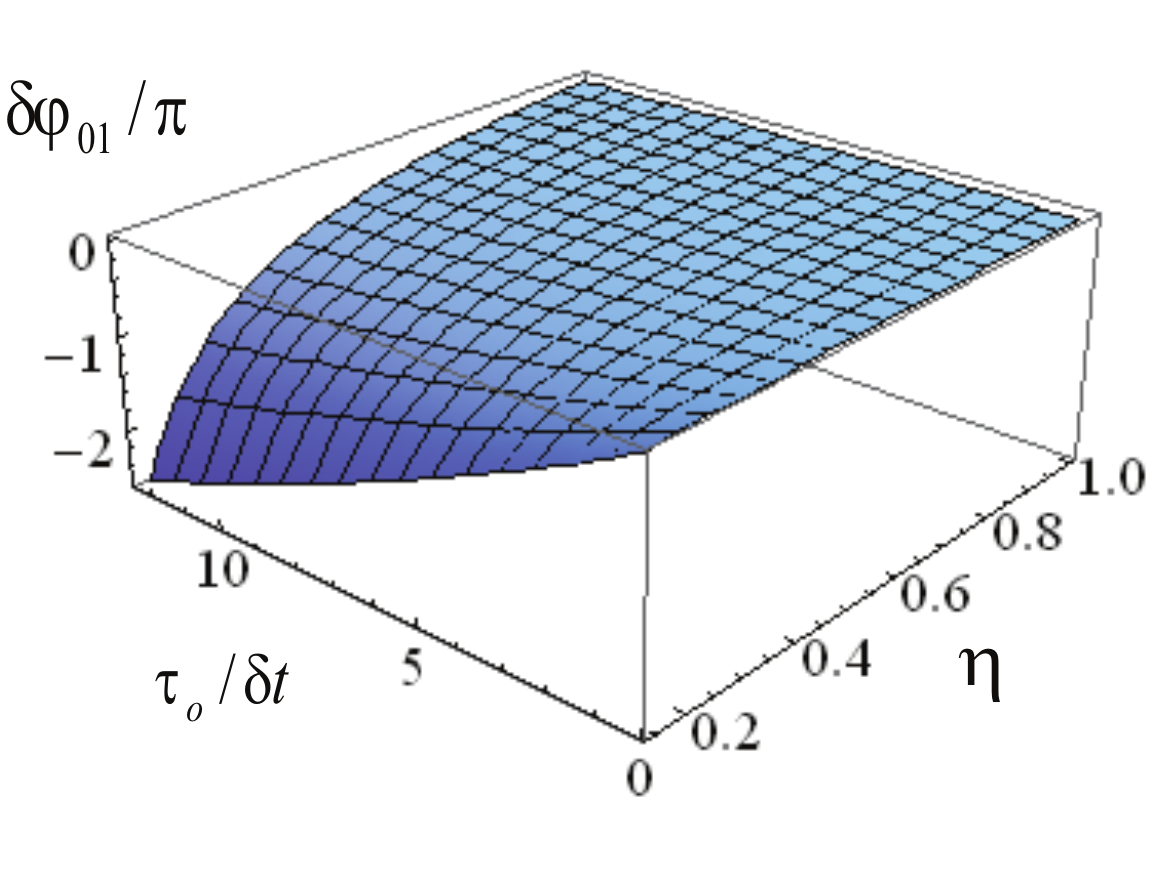}\\
  \caption{Phase shift $\delta \varphi_{01}$ for quantum decompression as a function of the time delay $\tau_o$ and the compression parameter $\eta$, with dephasing time $t_1=20\delta t$, $\Delta_{inh}\delta t=10$, and $2 \pi \zeta/\chi=6\pi$.}
  \label{total_phase_shift1decomp}
\end{figure}

Note that a given dephasing time $\tilde{t}$ determines close-to-homogeneously broadened slices of length $\delta l$ in the atomic medium with $\delta l\approx (\chi \tilde{t})^{-1}$ where coherence, hence radiation, remains. We believe that coupling between light and collective atomic coherence (as included in the analyzed equations) within these slices plays an important role for those frequency shifts, which are proportional to the number of atoms $n=N\delta l$ in each slice, and the square of the photon-atom coupling constant $g$ (i.e. $\delta\omega\propto g^2N\delta l$).
The nonlinear phase modulation also leads to an additional, possibly substantial phase difference between the two basic wavepackets after recall:
$
\delta\varphi_{01}=\int_{t_{echo}}^{t_{echo}+\tau_0/\eta} dt \delta\omega(t)
=\Phi(t_{echo}+\tau_0/\eta)-\Phi(t_{echo})$. The phase difference is depicted in Figs. \ref{total_phase_shift1comp} and \ref{total_phase_shift1decomp}. The effects of the non-linear phase shift are enhanced with increasing effective optical depth $\kappa_{eff}=2\pi\zeta/\chi$, time delay $\tau_o$ between the wave packets, and diminishes as the dephasing time $t_1$ increases, and the compression parameter $\eta$ approaches unity.

Phase and frequencies shifts in the here analyzed case of backward emission are a result of the presence of phase mismatch ($\delta k^{(l)}\neq 0$), or compression/decompression ($\eta\neq 1$), or both, but they do not occur in the phase-matched, reversible case. However, in the case of forward emission, the phase and frequency shifts also arise for non-compressed recall ($\eta =1$), provided the rephasing and dephasing times $\tilde{t}$ are bounded \cite{Hetet2008,Moiseev2008}.

Let us now discuss our figures of merit in the case of longitudinal broadening and compare it with transverse broadening. As mentioned above, the new basis states (Eqs. (\ref{temporal compressed wavepackets longitudinal})) do not transfer into each other through time translation alone, in contrast to the case of transverse broadening. For instance, assuming $\delta t$=100 ns, $\tau_o=200$ ns, $2\pi\zeta /\chi=6$, $\eta=3$ and a sufficiently large dephasing time $t_1/\delta t=20$, we find a frequency difference $\abs{\delta\omega_0-\delta\omega_1}\approx 2\pi\times$ 17.32 kHz . As this value is small compared to the spectral width of the wave packet $c\delta k \approx 2\pi\times$ 1.59  MHz, we can safely ignore this effect as a limitation to the fidelity. It is thus possible to map the new basis states onto the qubit basis states $\ket{0}$ and $\ket{1}$. Furthermore, as before, we can compensate for the additional relative phase arising during compression using an appropriate $\sigma_z$ rotation. Hence, the fidelity in longitudinal compression can be close to one.

The recall efficiency $\epsilon^{(l)}$  is given by
\bea
\label{recall efficiency longitudinal}
\epsilon^{(l)}=(R^{(l)})^2
=(\Gamma_{\gamma_{eg}}(t_1+t_{\delta k}^{(l)}))^2
\abs{M^{(l)}(\zeta/\chi,\eta )}^2\{\alpha ^2 e^{ -2 (1 +1/ \eta )\gamma _{eg} \tau_o }
+ \beta ^2 \}.
\eea
\noindent
As compared to Eq. (\ref{recall efficiency}), we find the same function $(\Gamma_{\gamma_{eg}} (t_1+t_{\delta k}^{(l)}))^2$, which characterized atomic decay during the storage process (Eq. (\ref{Gamma})), only with slightly modified argument (now including an additional delay $t^{(l)}_{\delta k}$). 
Note that phase mismatch only leads to a decrease of the efficiency when paired with significant atomic phase relaxation $\gamma_{eg}$, in contrast to the transverse case (Eq. (\ref{M})).

\begin{figure}
  \includegraphics[width=0.45\textwidth]{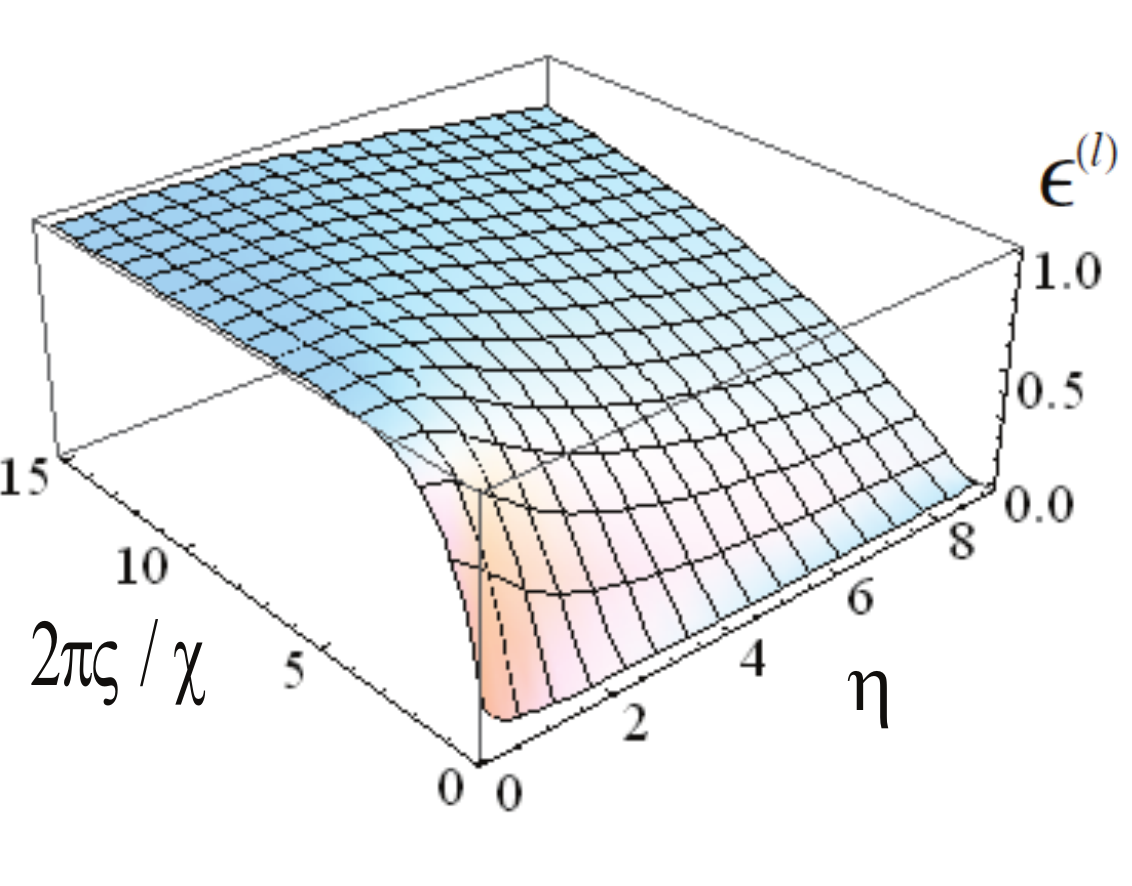}
  \caption{Recall efficiency for longitudinal broadening as a function of the  compression parameter $\eta$ and the effective optical depth  $2\pi\zeta/\chi$ (see Eq.((\ref{recall efficiency longitudinal})).}
    \label{figure_efficiency_longitudinal_scheme}
\end{figure}

\begin{figure}
  \includegraphics[width=0.45\textwidth]{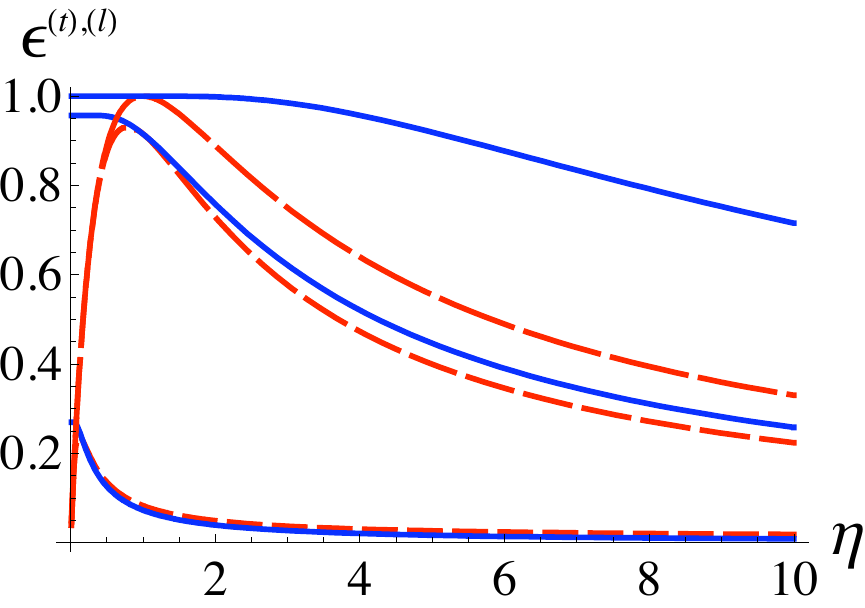}
  \caption{Recall efficiencies for compression schemes employing transverse (red-dashed lines) and longitudinal broadening (blue lines). The initial optical depths
  $\kappa_{eff}=\alpha_{o}L=2 \pi\zeta/\chi$ are $ 0.1 \pi,  \pi,  4 \pi$ (bottom to top sets of two curves). For small optical depth and sufficiently large compression factor $\eta$, the efficiencies $\epsilon^{(t)}(\eta)$ and $\epsilon^{(l)}(\eta)$ are equal, as described by Eq. (\ref{recall efficiency for small optical depth}). As the optical depth increases, schemes based on longitudinal broadening perform better.}
  \label{figure_efficiency_TL}
\end{figure}

Figure \ref{figure_efficiency_longitudinal_scheme} depicts the efficiency $\epsilon^{(l)}$ in the case of negligible phase relaxation (i.e. $\gamma_{eg}=0$) as a function of effective optical depth $2\pi\zeta/\chi$ and compression parameter $\eta$. Furthermore,
a comparison of efficiencies of the longitudinal and transverse compression schemes for various initial optical depths is shown in Fig. \ref{figure_efficiency_TL}. For small initial optical depth ($\kappa_{eff}=2 \pi\zeta/\chi\lesssim 0.5\pi$, or $\kappa_{eff}=\alpha_oL\lesssim 0.5\pi$, respectively) and large compression parameter ($\kappa_{eff}/\eta < 1$), both schemes feature the same  efficiency:
\bea
\label{recall efficiency for small optical depth}
\epsilon^{(l)}(\eta,\kappa_{eff})=\epsilon^{(t)}(\eta,\kappa_{eff})\cong\kappa_{eff}^2/\eta.
\eea

\noindent
This behavior is expected as reabsorption in transverse as well as longitudinally broadened media is negligible in the case of small optical depth. However, as the optical depth increases, reabsorption becomes more and more important in the case of transverse broadening, reflecting the departure from reversible light-atom interaction, while it is no issue in the case of longitudinal broadening, due to the correlation between atomic detuning $\Delta$ and position z (see Eq. (\ref{longitudinal broadening})). In particular, for longitudinal broadening, we find

\beq
\epsilon^{(l)}(\eta\rightarrow 0)= (1-\exp\{-\kappa_{eff}\})= 1|_{\kappa_{eff}\gg 1},
\eeq
\noindent
while decompression with transverse broadening yields a monotonously decreasing efficiency
\beq
\epsilon^{(t)}(\eta\rightarrow 0)= 4 \eta |_{\kappa_{eff}\gg 1}
\eeq
\noindent
and eventually becomes zero.

Furthermore, the compression efficiency $\epsilon^{(l)}$ is only limited by the optical depth $\kappa_{eff}$ during absorption, as well as by atomic relaxation. In particular, there is no compression factor dependent upper limit, in opposition to $\epsilon^{(t)}$ (see Eq. (\ref{efficiency fundamental limit transverse broadening})). Hence, provided sufficient optical depth $2\pi\zeta / \chi>1$ and $2\pi\zeta / (\eta\chi )>1$ (which may be difficult to achieve for large compression parameters), one can always achieve close-to-unit efficiency.  This gives longitudinal broadening a clear advantage over transverse broadening for quantum compression.

\begin{figure}
  \includegraphics[width=0.45\textwidth]{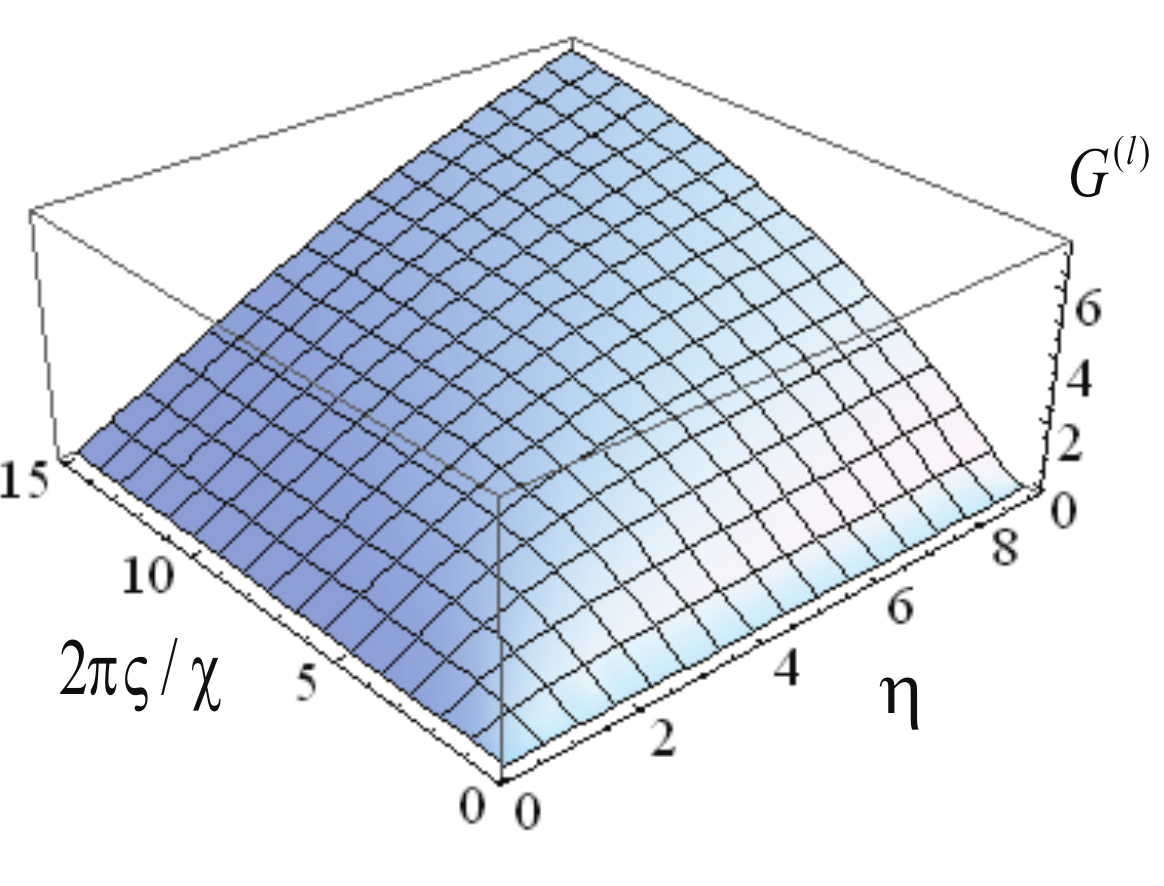}
  \caption{Gain for longitudinal broadening as function of compression parameter $\eta$ and effective optical depth $\kappa_{eff}=2 \pi\zeta/\chi$.}
  \label{Figure_gain_l}
\end{figure}

Obviously, the improved performance is reflected in the gain. As shown in Fig.  (\ref{Figure_gain_l}), we find that the gain can always be increased when increasing the effective optical depth or the compression parameter, as opposed to the case of transverse broadening where the gain was limited to four (see Eq. (\ref{max_gain_t})). Indeed, using Eqs. (\ref{gain_definition}) and (\ref{recall efficiency longitudinal}), and ignoring atomic relaxation (i.e. $\gamma_{eg}$=0), we find

\beq
G^{(l)}=\eta(1-\exp\{-\kappa_{eff})/\eta\}(1-\exp\{-\kappa_{eff}\}).
\eeq
\noindent
Hence, assuming sufficiently large effective optical depth, we find

\beq
G^{(l)}(\eta/\kappa_{eff}\rightarrow 0)=\eta  |_{\kappa_{eff}\gg 1}.
\eeq

\section{Feasibility study}

Before we conclude this article, let us briefly discuss possible implementations of our proposal. Quantum compression as considered here requires an atomic ensemble with (at least) three suitable energy levels. Two levels must be coupled resonantly by the photonic wavepacket to be compressed. The third level, generally another ground state, is employed for temporal mapping of the excited optical coherence onto ground state coherence using two counter-propagating $\pi$-pulses. This results in the implementation of the $2kz$ phase shift, i.e. in emission of light in backward direction \cite{Moiseev2001}. The width of the optical absorption line, which may be inhomogeneously broadened, should be sufficiently small. It should be possible to broaden the line in a controlled way via external fields, and the optical depth after broadening should be large. The maximum bandwidth of the optical wavepacket to be absorbed or re-emitted is given by the width of the externally broadened atomic transition, while the minimum bandwidth of the optical wavepacket is limited by the atomic linewidth before broadening (more precisely, the minumum optical bandwidth is a small multiple of the atomic linewidth as some controlled broadening is required for controlled dephasing or rephasing). These conditions are similar to those required for quantum state storage based on standard CRIB \cite{Tittel2009}, but without the need for a long storage time, which considerably relaxes material requirements.

Any material that fulfills these requirements would be a suitable candidate for quantum compression. We note that some aspects of the here presented theoretical study have recently been observed using atomic vapour, longitudinal broadening, and bright pulses of light \cite{Hosseini2009}. In the following, we will discuss RE-ion doped solid-state material in view of quantum compression.

RE doped solids have been studied extensively for data storage experiments based on stimulated photon echoes \cite{RE book}, and are currently being investigated for photon-echo quantum memory \cite{Tittel2009,Lvovsky2009}.
When implemented into crystals and cooled down to temperatures below 4 Kelvin, RE ions typically feature homogeneous linewidths (for optical transitions) of a few kHz, and values as small as 50 Hz have been reported \cite{Sun2002}. Linewidths in glasses are larger, but can still be around or below 1MHz \cite{Macfarlane2006,Staudt2006,Sun2006}. The transitions in crystals and glasses are inhomogeoeusly broadened, with values ranging from 40 MHz \cite{Macfarlane1992} in crystals to hundreds of GHz in crystals or glasses \cite{Sun2002,Zyskind1990}. The preparation of the initial, narrow absorption line therefore requires an initial preparation step via optical pumping \cite{Pryde2000,Nilsson2004,Crozatier2004,Lauritzen2008}. In low symmetry hosts, the RE ion's quantum states acquire permanent electric dipole moments, which allows shifting their energy levels through the application of a dc electric field -- the dc Stark shift (for a review on optical Stark spectroscopy of solids see \cite{Macfarlane2007}). A different dipole moment for two energy levels consequently results in a shift of the associated transition frequency, allowing, through the application of a field gradient, to broaden an isolated absorption line. Typical frequency shifts in RE ion doped materials are ten to hundred kHz for an electric field of one $Vcm^{-1}$.

Praseodymium doped Yttrium Orthosilicate (Pr$^{3+}$Y$_2$SiO$_5$) is a very promising candidate for the demonstration of quantum compression in RE solids. This crystal has been employed for storage of light pulses using longitudinal broadening \cite{Hetet2008}, and recall efficiencies up to 69\% have recently been reported \cite{Hedges2010}.Taking into account a spectral width of the $^3H_4\Leftrightarrow ^1D_2$ (605.9 nm wavelength) absorption line of 30 kHz after optical pumping \cite{Hetet2008}, and a Stark coefficient of 112.1kHz/(Vcm$^{-1}$) \cite{Graf1997}, we find that Pr$^{3+}$Y$_2$SiO$_5$ allows absorption and reemission of photonic wavepackets with bandwidths between  approx. hundred kHz and a few MHz, i.e. temporal durations between  approx. hundred nsec and a few $\mu$sec. This results in a maximum compression (scaling) factor $\eta$ of $\sim$100 (or $\eta\sim$1/100). Taking into account the demonstrated high efficiency storage \cite{Hedges2010}, one can expect a gain $G^{(l)}>10$. However, note that the maximum bandwidth for this transition is $\sim$10 MHz, due to limitations imposed by ground state hyperfine splitting on the preparation of the initial absorption line via optical pumping \cite{Nilsson2004}.

For RE materials with larger ground state splitting, allowing storage or recall of shorter wavepackets, it is interesting to consider RE-ion doped crystalline and amorphous waveguides. Due to the possibility to implement electrodes with spacing as small as $\sim$10 $\mu$m, large electric fields, i.e. large Stark shifts, can be obtained through application of modest voltages. So far, waveguides in Erbium or Thulium doped Lithiumniobate crystals and Erbium doped silicate fibres have been investigated in view of quantum state storage \cite{Staudt2006,Hastings-Simon2006,Staudt2007,Sinclair2009}, and Stark broadening up to $\sim$100 MHz seems feasible, allowing for storage or recall of pulses as short as $\sim$10 ns.

\section{Conclusion}

In summary, we have studied quantum compression and decompression of photonic time-bin qubit states
employing a generalized version of CRIB-based photon-echo quantum memory. Assuming high optical depth for storage and retrieval, we find, for the case of transverse broadening, that the recall efficiency  is limited by the compression factor $\eta$, while it reaches unity in the case of a longitudinal broadened medium. We also find, for transverse broadening, that the fidelity of recalled photonic time-bin qubits with the original qubit is one, regardless the compression factor, but that it is limited in the case of longitudinal broadening. Taking into account realistic experimental data, we foresee that quantum compression will be useful for quantum communication and computation applications. In particular, it enables enhancing the data rate in quantum communication schemes through temporal multiplexing, and allows mapping of broadband photons into small-bandwidth quantum memory.

Our analysis reveals new aspects of coherent photon-atom interaction, specifically photon-echo type interactions, and highlights the advantage of schemes employing longitudinal broadening over transverse broadening. The theory can be generalized in a straightforward way to encoding of quantum information into multiple ($n>2$) (discrete) temporal modes of photons, including compression of intense light fields (provided the number of photons is smaller then number of resonant atoms). However, quantum compression/decompression of light carrying quantum information with continuous spectrum (continuous quantum variables) requires further theoretical investigations that take into account added quantum noise in the case of limited recall efficiency. Finally, we note that pulse compression has been observed in the context of frequency conversion based on Raman adiabatic transfer of optical states \cite{Vewinger2007}, a protocol based on electromagnetically induced transparency that may provide an interesting alternative to our approach. 

\section*{Acknowledgments}
This work was supported by NSERC, iCORE (now part of Alberta Innovates - Technology Futures), GDC, as well as the Russian Foundation of Basic Research under grants No. 08-07-00449 and 10-02-01348-a, and Government contract of RosNauka 02.740.11.01.03.

\newpage
\subsection{Appendix -- Transverse broadening}
\label{AppendixA}

First, we find a solution of Eqs. (\ref{quantum evolution of field}) and (\ref{quantum evolution of atoms}) given an input photon amplitude $A_ {+} (t \leq t_1,z)$ and  excited atomic coherence $b_j (t\leq t_1)$. Assuming  all atoms to be initially in the ground state ($b_j (\tau=t_o\rightarrow -\infty)\rightarrow 0$) 
and using a temporal Laplace transformation for the amplitude $\bar {A}_{ip,+} (z) = \int_{t_o }^\infty
{d\tau } A_ {+}(t ,z)\exp \{ - p(t - t_o)\}$
we find the formal solution of  Eq. (\ref{quantum evolution of atoms})

\begin{equation}
\label{eq3}
b_j (t < t_1 ) = ig\frac{1}{2\pi }\exp \{ - i\omega _ + (t - t_o
)\} \nonumber\\
\int\limits_{t_o }^t {d t'} \int\limits_{ - \infty }^\infty
{d\nu } \bar {A}_{\nu , + } (z_j )\exp \{ - i\nu t ' - i(\omega _{eg} +
\Delta _j - \omega _ + )(t - t ') - i\delta \phi _{eg}^j (t ,t
')\},
\end{equation}

\noindent
where we have also used a backward Laplace transformation  $A_ + (t,z) = (2\pi )^{ - 1}\int_{ - \infty }^\infty {d\nu\bar{A}_{\nu, + } (z)\exp \{ - i\nu (t-t_o) \}}$ with $p=-i\nu$.
Putting Eq. (\ref {eq3}) in Eq. (\ref{quantum evolution of field}) and  taking into account phase relaxation as described in Eq. (\ref{homogeneous broadening}) we find the following expression for the sum of arbitrary atomic functions $F_j (t,z_j )$:

\[
\sum\limits_{j = 1}^N {F_j (t,z_j )\exp \{ - i\int_{t_o }^{t_1 } d t'\delta
\Delta _{eg}^j (t')\}\delta (z - z_j )\left| {_{N \gg 1 } } \right.}
\]

\begin{equation}
\label{defenition of summation}
 = (n_o S)\exp \{ - \gamma _{eg} (t_1 - t_o )\}\int\limits_{ - \infty
}^\infty {d\Delta } G(\Delta / \Delta_{inh} )F(\Delta ,t,z).
\end{equation}

\noindent
The Fourier components of the input field are given by

\begin{equation}
\label{eq2}
\bar {A}_{\nu , + } (z) = \exp \{ - \textstyle{1 \over 2}\alpha _ + (\nu
)z\}\bar {A}_{\nu , + } (0),
\end{equation}

\noindent
where 

\begin{equation}
\label{absorption coefficient s}
\alpha _+ (\nu) = \alpha_o(\gamma_{eg}
+\Delta_{inh})\int\limits_{ - \infty }^\infty
{d\Delta } \frac{G(\Delta / \Delta_{inh} )}{[\gamma _{eg} - i(\nu + \omega _ + -
\omega _{eg} - \Delta )]}
\end{equation}

\noindent
is the frequency-dependent absorption coefficient for an arbitrary, inhomogeneously broadened absorption line $G(\Delta / \Delta_{inh} )$ of the transverse type, and the on-resonant absorption coefficient $\alpha_o=2\pi n_o S g^2/[c(\gamma_{eg}+\Delta_{inh})]$.
The amplitude $A_ + (t>t_1,z=L)$  in Eq. (55) describes the light field at $z\geq L$ after interaction with the atomic medium. The Fourier component at the beginning of the medium $\bar {A}_{\nu , + } (0)$ is determined by the initial state of the input light at $z=0$.

We now consider inhomogeneous broadening with Lorentzian lineshape $G(\Delta /\Delta_{inh} ) = \textstyle{{\Delta_{inh} } \over {\pi (\Delta ^2 + \Delta_{inh}^2
)}}$, assuming the spectral width of the input field narrow compared to the atomic line width ($\delta \omega _ + \ll \Delta_{inh} $). This results in $\alpha _ + (\nu) = \frac{\alpha _o (\gamma _{eg} + \Delta_{inh}
)}{[\gamma _{eg} + \Delta_{inh} + i(\omega _{eg} - \omega _ + - \nu )]}$ in Eq. (\ref{absorption coefficient s}).

To allow for light re-emission in backward direction, we change the phases of the atomic amplitudes through the application of two short $\pi$ laser pulses that temporally map atomic coherence on the optically excited transition onto a ground state transition \cite{Moiseev2001,Moiseev2004}, as discussed in Sec. \ref{section symmetric CRIB}. In general, this phase matching operation leads to some phase mismatch between the atomic coherence and the irradiated echo signal that we quantify through an additional atomic phase shift $\delta k z_j$:
$b_j (t_1+(\textstyle{\epsilon\rightarrow 0}))=b_j (t_1)\exp\{i\delta k z_j\} $ in Eqs. (\ref{quantum evolution of field}) and (\ref{quantum evolution of atoms}) (with index $\sigma=-$ denoting echo signal emission and $\Delta_j(t>t_1)=-\eta\Delta_j$). Taking into account the initial atomic amplitudes $b_j (t)=b_j (t_1)\exp\{i\delta k z_j+i\eta\Delta_j(t-t_1)-i\delta \phi _{eg}^j(t,t_1)\}$  and Eq.
(\ref{defenition of summation}), we find the irradiated light field to be described by:

\[
\textstyle{\partial \over {\partial z}}\bar {A}_ {ip, -} (z) =
\]

\begin{equation}
\label{echo field equation}
 = (\pi n_o Sg^2 / c)\bigg ( \int\limits_{t_1 }^\infty {d\tilde {\tau }}
\exp \{ - p\tilde {\tau }\}\bar {P}_{echo} (\tilde {\tau },z)
 +\frac{\bar {A}_{ip, -} (z)}{p + (\gamma_{eg} + \eta
\Delta_{inh} ) + i(\omega _{eg} - \omega _ - )}\bigg ),
\end{equation}

\noindent
where $\bar {P}_{echo} (\tilde {\tau },z)$ is the rephased atomic coherence that serves as a source
for the irradiated field:
\[
\bar {P}_{echo} (\tilde {\tau },z) = \exp \{ - (1 + \eta )\gamma _{eg}
(\tilde {\tau } - t_1 )\}
\]

\[
\exp \{i[\delta k z+\omega _ - - \omega _{eg}  + \eta (\omega _ + - \omega _{eg}
)]\tilde {\tau } + i(\omega _{eg} - \omega _ + )(1 + \eta )t_1 \}\}
\]

\begin{equation}
\label{echo polarization}
\int\limits_{ - \infty }^\infty {d\nu } \frac{2\Delta_{inh} \exp \{i\nu[\eta
(\tilde {\tau } - t_1 ) - t_1 ]\}}{[(\omega _ + + \nu - \omega _{eg} +
i\gamma _{eg} )^2 + \Delta_{inh}^2 ]}\exp \{ - \frac{\alpha _o z}{2[1 +
i(\omega _{eg} - \omega _ + - \nu ) / (\gamma _{eg} + \Delta_{inh} )]}\}
\bar {A}_{\nu , + } (0).
\end{equation}

\noindent
We used again the temporal Laplace  transformation for light and atom amplitudes within the temporal interval [$t_1, \infty$], new variables $\tilde\tau=t+z/c$, $z=z$, and we have taken into account that initially $A_-(t_1,0<z<L)=0$. The solution of Eq. (57) has the form of a double integral

\[
A_-(\tilde\tau ,z) =
\]

\[
\frac{i}{2\pi^2}\int\limits_{ - \infty }^\infty {d\nu } \frac{\Delta_{inh}
\bar {A}_{\nu , + } (0)\exp \{i(\varphi-\nu t_1)\}}
{[(\omega _ + - \omega _o + \nu + i\gamma _{eg} )^2 + \Delta_{inh}^2 ]}
\int\limits_{ - \infty }^\infty {d\Omega }
\frac{\exp \{i\Omega (\tilde{\tau } - \tilde {\tau }_o )\}}
{[\alpha _ + (\nu)+ \alpha _ - (\Omega)  -
2i\delta k]}
\]

\begin{equation}
\label{complete solution}
\frac{\alpha _o (\gamma _{eg} + \Delta_{inh} )}{[\Omega - \Omega _1 (\nu)]}
\{\exp [ - \textstyle{1 \over 2}\alpha _ + (\nu)z + i\delta kz] - \exp [ -
\textstyle{1 \over 2}\alpha _ + (\nu)L + \textstyle{1 \over 2}\alpha _ -
(\Omega )(z - L) + i\delta kL]\},
\end{equation}

\noindent
where $\omega _ - = \omega _{eg} + \eta (\omega _{eg} -
\omega _ + )$ is the carrier frequency of the echo field,
$\varphi
=  (1+\eta)(\omega _{eg} - \omega _+) t_1,$
$\Omega _1 (\nu) = (\omega _ - - \omega _{eg} ) + \eta (\omega _ + - \omega
_{eg} + \nu) + i(1 + \eta )\gamma _{eg}$, and 
 $\alpha _ - (\nu) = \frac{\alpha _o (\gamma _{eg} + \Delta_{inh} )}{[\gamma
_{eg} + \eta \Delta_{inh} + i(\omega _{eg} - \omega _ - + \nu)]}$ is the
absorption coefficient of the atomic system during echo emission. It is  characterized by the compressed inhomogeneous broadening $\eta\Delta_{inh}$.  Assuming again the  inhomogeneous broadening to be large compared with the initial light spectrum, $\Delta_{inh} \gg \delta \omega _f $, we can simplify the dependence on the phase mismatch factor:

\begin{equation}
\label{eq77}
\frac{2}{\alpha _ + (\nu)+ \alpha _ - (\Omega) - 2i\delta k} \cong
\frac{2}{\alpha _o (\gamma _{eg} + \Delta_{inh} )}\frac{[\gamma _{eg}
+ \chi \Delta_{inh} + i(\omega_{eg} - \omega _ -+\Omega)][\gamma _{eg} + \Delta_{inh} +
i(\omega _{eg} - \omega _ + - \nu )]}{[2\gamma _{eg}+(1 + \chi )\Delta_{inh}
+i(2\omega _{eg} - \omega _ + -
\omega _ - - \nu+\Omega)- 2i\delta
k\frac{(\gamma _{eg} + \chi \Delta_{inh} )}{\alpha _o }]}.
\end{equation}

The main part of the integration in Eq. (\ref{complete solution}) over $\Omega $ in the complex plane  is determined
 by the pole singularity at $\Omega =\Omega _1 (\nu )$, while the other singularities describe fast decaying signals associated with the large inhomogeneous broadening. After the integration we obtain the following echo field amplitude at the medium output ($z=0$):

\[
A_ - (\tilde {\tau },z = 0) =
\frac{\eta \exp \{i\varphi - (1 + \eta)\gamma _{eg} (\tilde {\tau } - t_1 )\}}{\pi \left[ {(\eta +
1) + (1 - \eta )\gamma _{eg} / \Delta_{inh} } \right]}
\]

\begin{equation}
\label{first solution}
\int\limits_{ - \infty }^\infty {d\nu } \frac{\exp \{i\nu[\eta (\tilde {\tau }
- t_1 ) - t_1 ]\}\{1 - \exp [ - \textstyle{1 \over 2}\alpha (\nu,\eta )L +
i\delta kL]\}}{\left\{ {1 + i\frac{(\eta - 1)(\nu + \omega _ + - \omega _{eg}
) / \Delta_{inh} }{(\eta + 1) + (1 - \eta )\gamma _{eg} / \Delta_{inh} } -
2i\frac{(\delta k / \alpha _o )(\eta + \gamma _{eg} / \Delta_{inh} )}{(\eta +
1) + (1 - \eta )\gamma _{eg} / \Delta_{inh} }} \right\}}\bar {A}_{ + ,\nu} (0),
\end{equation}

\noindent
where  $\alpha (\nu,\eta ) = [\alpha _ + (\nu) + \alpha _ - (\nu)]$ describes the combined influence of absorption and dispersion effects on the echo field.

Eq. (\ref{first solution}) allows calculating the compression efficiency. For instance, setting $\eta=1$, we can derive the echo field in the presence of phase mismatch \cite{Moiseev2004,Moiseev2004b}, which limits the recall efficiency. Setting $\delta k=0$, we find the echo to be a perfect, temporally reversed copy of the input field, provided the optical depth is sufficiently large for all spectral component of the field ($\alpha (\nu,\eta )L\gg 1$. This confirms previously obtained results \cite{Moiseev2004,Sangouard2007,Kraus2006}.  For sufficiently large inhomogeneous broadening $\delta \omega _f/\Delta_{inh} \gg 1$ we find $\alpha (\nu,\eta ) \cong (1 + 1 / \eta )\alpha _o $ and Eq. (\ref{first solution}) can be simplified to:

\[
A_ - (\tilde {\tau },z = 0) =
\]

\begin{equation}
\label{second solution}
\epsilon_o(\eta)^{1/2}
M^{(t)}(\delta k, \alpha _oL)
\exp \{ i\varphi - (1 + \eta)\gamma _{eg} (\tilde {\tau } - t_1 )\}
\sqrt{\eta}
A_+[-\eta (\tilde {\tau }- t_1 ) + t_1 ],
\end{equation}

\noindent
where $M^{(t)}(\delta k, \alpha _oL)$ and $\epsilon_o^{1/2}(\eta)$ are given in
Eqs. (\ref{M}) and (\ref{epsilon_0}). While we are only interested in the quantum state of the retrieved and compressed photonic qubit (as described by Eq. (\ref{final solution for Gaussian qubits transverse})), which we obtain through Eq. (\ref{second solution}) assuming initial photonic time-bin qubits (Eq. (\ref{initial envelope})) with Gaussian shape of the basis wave packets, we note that one can also derive the amplitude of the atomic state (Eq. (\ref{general atomic state})) after the echo signal emission by inserting Eq. (\ref{complete solution}) into Eq. (\ref{quantum evolution of atoms}).

\subsection{Appendix - Longitudinal broadening}
\label{AppendixB}

As in the above discussed case of transverse broadening, we are only interested in the retrieved photonic time-bin qubit state. Averaging Eq. (\ref{quantum evolution of atoms}) over the phase fluctuations, and after a variable transformation to a moving reference frame  $\tau = t - z / c$, $z = z$, the light-atom equations for $t < t_{1}$ become

\begin{equation}
\label{long field equation }
\textstyle{\partial \over {\partial z}}A_ {1}(\tau ,z) = i(\pi {n_o S}
g^\ast / c) b_{o}(\tau ,z),
\end{equation}

\begin{equation}
\label{long atomic equation}
\textstyle{\partial \over {\partial \tau }}b_{o} (\tau,z ) =(i\chi z -
\gamma _{eg} ) b_{o} (\tau,z ) +
igA_ {1} (\tau ,z ),
\end{equation}

\noindent
where we used the substitution
$ b_j (\tau)= b_o(\tau,z_j)\exp \{ - i\omega_ {21}\tau\}$ with $b_o(\tau,z)$ describing averaged local atomic coherence, and
$A_1 (\tau ,z) = A_ + (\tau ,z)\exp \{ - i(\omega_ + - \omega _{eg} )\tau \}$.
The field amplitude $A_ + (\tau ,z)$\textsf{ } is given by Eq. (\ref{initial envelope}) with additional factor
$\exp \{i\phi _{lg} \}$ where $\phi _{lg} = \textstyle{1 \over 2}\omega _ + L / c$.

\noindent
The general solution of
Eqs. (\ref{long field equation }) and (\ref{long atomic equation}) is obtained using a temporal Laplace transformation \cite{Moiseev2008}. We are interested only in the atomic coherence  excited by the input light, which gives rise to the echo field irradiated after further evolution, while the initial light field disappears in the medium ($A_1 (\tau\gg\delta t,z) \to 0)$:

\begin{equation}
\label{eq59}
b_{o}(\tau_1 ,z) = - \frac{i}{2\pi }\int\limits_{ - \infty }^\infty {d\omega e^{ -
i\omega \tau_1 }} \frac{(\omega - \chi L / 2 + i\gamma _{eg} )^{i\zeta / \chi
}}{(\omega + \chi z + i\gamma _{eg} )^{1 + i\zeta / \chi }}\tilde {A}_1
(\omega , - L / 2).
\end{equation}

\noindent
$\tilde {A}_1 (\omega , - L / 2) = \int_{ - \infty }^\infty {d\tau
e^{i\omega \tau }} A_1 (\tau ; - L / 2)$ is the field spectrum at the input of the atomic medium ($z=-L/2$), and
$\zeta = \pi n_o Sg^2 / c$. Similar to the transverse scheme, we find the
atomic coherence on the resonant optical transition $1\rightarrow2$ at the moment of re-emission of light using the phase mismatch factor $\delta k z$: $b_{o}(\tau _1+(\textstyle{\epsilon\rightarrow 0}))=b_{o}(\tau _1)\exp\{i\delta k z_j\}.$

We now analyze the backward echo emission. After changing the atomic
detunings $\Delta (\tau > \tau _1 ) = \chi 'z$, we find the
following system of equations for the atomic coherence and irradiated echo field
$A_2 (\tilde\tau >\tau _1,z)$ (where $\tilde\tau=t+z/c$):

\begin{equation}
\label{2long amplitude equation }
\textstyle{\partial \over {\partial z}}A_ {2}(\tilde\tau ,z) = -i(\pi {n_o S}
g^\ast / c) b_{o}(\tilde\tau ,z),
\end{equation}

\begin{equation}
\label{2long atomic equation}
\textstyle{\partial \over {\partial \tilde\tau }}b_{o} (\tilde\tau,z ) =-(i\chi z +
\gamma _{eg} ) b_{o} (\tilde\tau,z ) +
igA_ {2} (\tilde\tau ,z )
\end{equation}
\noindent
where
$A_2 (\tau ,z) = A_ - (\tau ,z)\exp \{ - i(\omega_ - - \omega _{eg} )\tilde\tau \}$, and
$\omega_-$ is the new carrier frequency.
Using a Laplace transformation in the solutions of Eqs. (\ref{2long amplitude equation })
and (\ref{2long atomic equation}),
we find $A_2 (\tilde\tau ,z<L/2)$ as a function of the
stored atomic coherence $b_{o}(\tilde {\tau _1},z)$:

\[
A_2 (\tau ,z = - L / 2) = \frac{\zeta }{(2\pi )^2}
\int\limits_{ - \infty }^\infty {d\omega e^{ - i\omega (\tau -
t _1 )}} \int\limits_{ - L / 2}^{L / 2} {dz'} \frac{( - \chi 'L / 2 -
\omega - i\gamma _{eg} )^{ - i\frac{\zeta }{\chi '}}}{(\chi 'z' - \omega -
i\gamma _{eg} )^{1 - i\frac{\zeta }{\chi '}}}\exp \{i\delta kz'\}
\]

\begin{equation}
\label{1 echo field in long scheme}
\int\limits_{ - \infty }^\infty {d\omega 'e^{ - i\omega t_1}}
\frac{(\omega ' - \chi L / 2 + i\gamma _{eg} )^{i\zeta / \chi }}{(\omega
' + \chi z' + i\gamma _{eg} )^{1 + i\zeta / \chi }}\tilde {A}_1 (\omega
',-L/ 2 ),
\end{equation}

\noindent
where we used $\tau_1=t_1$. Taking into account backward emission of the echo field, as well as large inhomogeneous broadening with respect to the input light spectral width, and using the simplifications

\[
(\omega ' - \chi L / 2 + i\gamma _{eg} )^{i\zeta / \chi }
\cong (\chi L / 2)^{i\zeta / \chi }\exp \{\gamma _{eg} \tau _m - \pi \zeta
/ \chi \}\exp \{ - i\tau _m \omega '\},
\]

\[
( - \chi 'L / 2 - \omega - i\gamma _{eg} )^{ - i\zeta / \chi '}
\cong (\chi 'L /2)^{ - i\zeta / \chi '}\exp \{\gamma _{eg} \tau _{m} ' - \pi \zeta / \chi
'\}\exp \{ - i\tau _{m} ' \omega \},
\]

\noindent
where $\tau _{m} ' = (\zeta / \chi ')\frac{(\chi 'L / 2)}{[(\chi 'L / 2)^2 +
\gamma _{eg}^2 ]} \cong \frac{(\zeta / \chi ')}{(\chi 'L / 2)}$, and $\tau _m =
(\zeta / \chi )\frac{(\chi L / 2)}{[(\chi L / 2)^2 + \gamma _{eg}^2 ]} \cong
\frac{(\zeta / \chi )}{(\chi L / 2)}$, we find, after changing the order of the integrations in Eq. (\ref{1 echo field in long scheme}) from $\omega, z', \omega'$ to
$\omega', z', \omega$:

\[
A_2 (\tau ,z = - L / 2) = \frac{\zeta }{(2\pi )^2}\exp \{\gamma _{eg} (\tau
_{m} ' + \tau _m ) - \pi \zeta (1 / \chi ' + 1 / \chi )\}(\chi 'L / 2)^{ -
i\zeta / \chi '}(\chi L / 2)^{i\zeta / \chi }
\]

\[
\int\limits_{ - \infty }^\infty {d\omega '\exp \{ - i\omega '(t_1+ \tau _m )\}} \tilde {A}_1 (\omega ', - L / 2)
\]

\begin{equation}
\label{2 echo field in long scheme}
\int\limits_{ - L / 2}^{L / 2} {dz'} \frac{\exp \{i\delta kz'\}}{(\omega ' +
\chi z' + i\gamma _{eg} )^{1 + i\zeta / \chi }}
\int\limits_{ - \infty }^\infty {d\omega } \frac{\exp \{ - i\omega (\tau -
t_1 + \tau _{m} ' )\}}{(\chi 'z' - \omega - i\gamma _{eg} )^{1 - i\zeta /
\chi '}}.
\end{equation}

\noindent
Taking again into account that $\chi L \gg \delta \omega_f$, and using the
tabled integral 

\[
(\chi 'L / 2)^{ - i\zeta / \chi '}\int\limits_{ - \infty
}^\infty {du} \frac{\exp \{iuT\}}{(u - i\gamma _{eg} )^{1 - i\zeta / \chi
'}} =
\]

\[
-\frac{2\pi \eta _\chi (T)}{ \left( {\zeta / \chi '} \right)\Gamma [ - i\zeta / \chi
']}\left( {T \chi 'L / 2} \right)^{ - i\zeta / \chi '}\exp \{ - \gamma
_{eg} T + \textstyle{1 \over 2}\pi \zeta / \chi '\},
\]
\noindent
where $\eta _\chi (T \ge 0) = 1$ and $\eta _\chi (T < 0) = 0$ is a Heaviside
function, we integrate Eq.(\ref{2 echo field in long scheme}) over $\omega $, then over z'  and $\omega '$, leading to

\[
A_2 (\tau ,z = - L / 2) = - \frac{2\pi \chi '}{\zeta }\frac{\exp \{ -
\textstyle{1 \over 2}\pi \zeta / \chi ' - \textstyle{1 \over 2}\pi \zeta /
\chi \}}{\Gamma [ - i\zeta / \chi ']\Gamma [i\zeta / \chi ]}
\eta _\chi (\tau - t _1 + \tau _m - {\delta k}/{\chi '})
\]

\[
\left( {(\chi 'L / 2)(\tau - t _1 + \tau _{m} ' )} \right)^{ - i\zeta /
\chi '}\mbox{ }\left( {(\chi 'L / 2)\vert \tau - t _1 + \tau _{m} ' -
{\delta k}/{\chi '}\vert } \right)^{i\zeta / \chi }
\]

\begin{equation}
\label{3 echo field in long scheme}
\exp \{ - \gamma _{eg} (1 + \eta)(\tau - t_1 ) + \gamma
_{eg} \eta\tau _z \}
A_1 \{ - \eta[\tau - (1 + 1/ \eta)t_1 - \tau _z ]\},
\end{equation}

\noindent
where $\eta=\chi '/ \chi$ is the compression factor, and  $\Gamma [ \pm i x]$ are Gamma functions.
Interestingly, as described by Eq. (\ref{3 echo field in long scheme}), phase mismatch does not effect to the quantum efficiency of the echo emission but only leads to a temporal shift of the echo emission to $\tau' _{echo} + \tau _z$ (where $\tau' _{echo} = ( 1+1/\eta) t _1 $,
$\tau _z = \delta k/\chi ' + \tau _m/ \eta - \tau_{m} ' $).
However, this shift results in an additional nonlinear phase shift due to the deviation from perfect temporal reversibility. Furthermore, we find that the echo emission is conditioned on a phase mismatch 
$\delta k/\chi ' > \tau _{m} ' + \delta t - (t _1 +
\tau _m )/\eta$. Taking into account that $\tau _m \leq \delta t$ and
$\tau _{m} ' \leq \delta t$, we find $
 \delta k /\chi  > -  t_1$.

Finally taking into account the relations between $A_2 (\tilde\tau,z)$ and $A_- (\tilde\tau,z)$, and
between $A_1 (\tilde\tau,z)$ and $A_+ (\tilde\tau,z)$, we find the new
carrier frequency
$\omega_-=\omega_{eg}+\eta (\omega_{eg}-\omega_+)$ of the echo signal, which coincides with the case of  transverse broadening. We also find
$A_-(\tilde\tau,z )=
A_2 (\tilde\tau,z )\exp\{ i\eta(\omega_{eg}-\omega_+)\tilde\tau\}$. Using this relation, together with
$A_1 \{ - \eta(\tau - \tau'_{echo} - \tau _z )\}=
A_+ \{ - \eta(\tau - \tau'_{echo} - \tau _z )\}
\exp\{i\eta(\omega_+-\omega_{eg})(\tau - \tau'_{echo} - \tau _z )\}$ in Eq.
(\ref{3 echo field in long scheme}), we find the solution for $A_-(\tilde\tau,z )$ as a function of $A_+$. Putting the obtained solution into Eq.(\ref{general photonic wavepacket}), assuming again a Gaussian shape of the photonic wave packets, we obtain the solution for the retrieved photonic time-bin qubit state given in Eq. (\ref{final solution for Gaussian qubits longitudinal}) after a simple algebraic calculation.

\end{document}